\documentclass[twocolumn,floatfix]{revtex4}

\usepackage{amsmath}
\usepackage{graphicx}
\usepackage{amssymb}

\def\rootfig{}

\begin{document}

\title{Phase Separation and Dynamics of Two-component Bose-Einstein Condensates}

\author{
R. Navarro$^1$,
R. Carretero-Gonz\'alez$^{1,}$\footnote{URL: {\tt http://rohan.sdsu.edu/$\sim$rcarrete/}},
P.G. Kevrekidis$^2$}
\affiliation{$^1$Nonlinear Dynamical Systems Group%
\footnote{URL: {\tt http://nlds.sdsu.edu}}, Computational Science
Research Center, Department of Mathematics and Statistics, San Diego
State University, San Diego CA, 92182-7720, USA\\
$^2$Department of Mathematics and
Statistics, University of Massachusetts, Amherst, MA 01003-4515 USA}

\begin{abstract}
We study the interactions between two atomic species in a binary
Bose-Einstein condensate to revisit the conditions for miscibility,
oscillatory dynamics between the species, steady state solutions and their
stability.
By employing a variational approach for a quasi
one-dimensional, two-atomic species, condensate we
obtain equations of motion for the
parameters of each species: amplitude, width, position and phase.
A further simplification leads to a reduction of the dynamics
into a simple classical
Newtonian system where components oscillate in an effective
potential with a frequency that
depends on the harmonic trap strength and the interspecies
coupling parameter.
We develop
explicit conditions for miscibility that can be interpreted as a
phase diagram that depends on the harmonic trap's strength and
the interspecies species coupling parameter.
We numerically illustrate the bifurcation scenario whereby
non-topological, phase-separated states of increasing complexity
emerge out of a symmetric state, as the interspecies coupling
is increased. The symmetry-breaking
dynamical evolution of some of these states is numerically monitored
and the associated asymmetric states are also explored.
%
\end{abstract}

\date{Submitted to {\em Phys. Rev. A}, 2009}


\maketitle

\section{introduction}
Over the past 15 years, the study of Bose-Einstein condensates (BECs)
has gained a tremendous momentum, stemming from an intense
and wide variety of theoretical, as well as experimental
studies that have now been summarized in a number of books
\cite{Pethick-book,stringari}. One of the particularly
intriguing aspects of the system is its effective nonlinearity
stemming from a mean-field representation of the
inter-atomic interactions. This, in turn, has led to a wide
range of developments in the area of nonlinear matter waves
in BECs \cite{Our-book} and the drawing of natural parallels between this
field and that of nonlinear optics, where similar nonlinear
Schr{\"o}dinger (NLS) types of models arise \cite{kivshar}.

One of the particularly interesting aspects of investigation of
BECs within the realm of NLS (often referred to in the BEC
context as Gross-Pitaevskii) equations is based on the
examination of multi-component systems. Starting from the
early work on ground-state solutions
\cite{shenoy,esry} and small-amplitude excitations \cite{excit} of the order
parameters, numerous investigations have been focused on
the study of two hyperfine states or two different atomic
species that can be condensed and confined concurrently.
More specifically, a few among the numerous topics investigated
involved the structure and dynamics of binary
BECs \cite{Marek,boris2,boris3},
the formation of domain walls between immiscible species \cite{Marek,healt},
bound states of dark-dark \cite{obsantos}, or dark-bright
\cite{anglin,sengstock}, or coupled-vortex \cite{ueda_review},
or even spatially periodic states \cite{decon}. The early
experimental efforts produced such binary mixtures of
different hyperfine states of $^{87}$Rb \cite{myatt} and of
$^{23}$Na \cite{nake}, but also of
mixed condensates \cite{dsh}.  Efforts
were later made to create two-component BECs with different atomic species,
such as $^{41}$K--$^{87}$Rb \cite{KRb} and $^{7}$Li--$^{133}$Cs \cite{LiCs},
among others. Recently the interest in multi-component BECs has
been renewed by more detailed and more controlled experimental
results illustrating the effects of phase separation
\cite{cornell,usdsh,wieman}, which have, in turn, motivated
further theoretical studies in the subject \cite{tsubota,karali};
see also \cite{usreview} for a recent review.

Although in
the present work, we will focus on two-component condensates,
it is relevant to note in passing the increasingly growing
interest in three-component, so-called, spinor condensates \cite{cahn}.
Among the numerous themes of investigation within the
latter context we mention spin domains \cite{Stenger1998a},
polarized states \cite{spindw},
spin textures \cite{spintext}, as well as
multi-component (vectorial) solitons of
bright \cite{wad1a,wad1b,boris,zh}, dark \cite{wad2}, gap \cite{ofyspin}, and
bright-dark \cite{ourspin} types.

Our aim in the present manuscript is to revisit the theme of
binary condensates in quasi-one-dimensional BEC settings, in an
attempt to offer additional both analytical and numerical insights
on the phenomenology of phase separation.
Our manuscript is organized as follows.  In
Sec.~\ref{sec:VA}, a Gaussian trial function is used in a
variational approach to obtain six first-order ordinary differential
equations (ODEs) for the time evolution of the parameters of the
two-component
ansatz: position, amplitude, width, phase, wave number and chirp. In
Sec.~\ref{ssec:bif}, the fixed points of the system of ODEs are
obtained to yield the equilibrium position, amplitude, and width of
the two species. Bifurcation diagrams of the ansatz' parameters are
produced as a function of the interspecies coupling strength. In
Sec.~\ref{ssec:phas}, phase diagrams are produced and analytical
conditions for
miscibility are obtained relating the  interspecies coupling
with the system's chemical potential and parabolic trap strength.
The dynamics of this system is compared to
results obtained by numerical integration of the Gross-Pitaevskii
equation in Sec.~\ref{ssec:dyn}. In Sec.~\ref{ssec:newt} the
system of ODEs is further reduced, upon suitable approximations,
to a classical Newtonian system; the latter is simple to analyze and
instructive with respect to the interpretation of the fundamental
interactions driving the system's dynamics.
In Sec.~\ref{sec:ES}, we numerically analyze in a systematic way the
existence and stability of higher excited phase-separated states as a
function of the interspecies interaction.
The dynamical instability evolution of the latter class of states
is examined numerically in Sec.~\ref{sec:dynamics}.
Motivated by the numerical experiments of Sec.~\ref{sec:dynamics},
in Sec.~\ref{sec:asym} we study the existence of asymmetric states 
when the chemical potentials of the two components differ.
Finally, in Sec.~\ref{sec:conclu} we summarize our
results and present some interesting directions for future
investigation.

\section{variational model\label{sec:VA}}
\subsection{Coupled equations\label{ssec:coup}}
The Gross-Pitaevskii (GP) equation, which is a variant of the NLS
equation accounting for the potential confining the atomic species,
governs the dynamics of
bosonic particles near absolute zero
temperatures \cite{Pethick-book,stringari}.
In the context considered herein (related to the case of
$^{87}$Rb which is common in relevant experiments \cite{myatt,usdsh}),
the two hyperfine states
of the same atom are described by a
set of coupled GP equations \cite{boris2}
\begin{eqnarray}
i\hbar\frac{\partial\psi_1}{\partial t}&=&\left(
-\frac{\hbar^2}{2m}\nabla^{2}+V_1 +g_{11}\left|\psi_1\right|^{2}+
g_{12}\left|\psi_2\right|^{2}\right)\psi_1\nonumber,\\
i\hbar\frac{\partial\psi_2}{\partial t}&=&\left(
-\frac{\hbar^2}{2m}\nabla^{2}+V_2 +g_{22}\left|\psi_2\right|^{2}+
g_{21}\left|\psi_1\right|^{2}\right)\psi_2\nonumber,\\
V_j&=&\frac{1}{2}\left(\omega_{jx}^2 x^2 +\omega_{jy}^2 y^2
+\omega_{jz}^2 z^2\right),j=1,2.\label{GPE_3d}
\end{eqnarray}
Here, $g_{jk}=4\pi\hbar^2a_ {jk}/m$ are the self coupling
interaction parameters of the first species for $j=1$, $k=1$, for
the second species $j=2$, $k=2$ and
for $j=1$, $k=2$ or $j=2$, $k=1$ are the cross species coupling
parameters. These
interaction strengths depend on the scattering lengths between same
($a_{11}$ and $a_{22}$) and different species ($a_{12}=a_{21}$).
%
The external harmonic trapping potential for each atomic
species, $V_j=V_j(\textrm{{\bf r}})$ depends on the radial distance
$\textrm{{\bf r}}$ from the center of the trap, while  the atomic density
is given by the square modulus of
$\psi_j=\psi_j(\textrm{{\bf r}},t)$.
The atomic mass  is denoted by $m$.

We will make the customary assumption
that the external trap's effect on each
species is the same, making $V_1(\textrm{{\bf r}})=V_2(\textrm{{\bf
r}})$. More importantly, in the interest of analytical
tractability of our results, we will also assume that the self
interactions for each species is the same, $g_{11}=g_{22}$.
Both assumptions are very good approximations of the physical
reality,
although they are not exact; see e.g., the relevant discussion
of \cite{usdsh}.

In a highly anisotropic trap, where the frequency of the
longitudinal component of the trap is much smaller than the
transverse components $\omega_x\ll\omega_y=\omega_z$, an effective
one-dimensional (1D) system of partial differential
equations (PDEs) can be obtained \cite{Our-book}
\begin{eqnarray}
i\frac{\partial u_1}{\partial t} &=&\left(-\frac{1}{2}\frac{\partial
^2 }{\partial
x^2}+\frac{\Omega^2}{2}x^2+\left|u_1\right|^{2}+g\left|u_2\right|^{2}\right)u_1,\label{GPE1_1d}\\
i\frac{\partial u_2}{\partial t} &=&\left(-\frac{1}{2}\frac{\partial
^2 }{\partial
x^2}+\frac{\Omega^2}{2}x^2+\left|u_2\right|^{2}+g\left|u_1\right|^{2}\right)u_2,\label{GPE2_1d}
\end{eqnarray}
where time, space, and wave function have  been rescaled to reduce
the system's parameters to just two ($\Omega$ and $g$).
In this 1D reduction, the chemical
potential corresponds to $\mu_{1D}$, $\psi_j\rightarrow u_j\sqrt{
g_{jj}/\mu_{1D}}$, $x\rightarrow x \sqrt{m\mu_{1D}}/\hbar$ is the
longitudinal distance from the center of the trap, $t\rightarrow t
\mu_{1D}/\hbar $ is the rescaled time,
$g=g_{12}/g_{11}=g_{21}/g_{22}$ is the rescaled species interaction
term.  In the literature, the condition of miscibility
$\Delta=(g_{12}g_{21}-g_{11}g_{22})/g_{11}=g^2-1$
is often used see e.g. \cite{Marek,boris2,boris3}.
The rescaled harmonic trap frequency is given by
$\Omega=\hbar\omega_x/\mu_{1D}$.
\subsection{Ansatz and Euler-Lagrange Equations\label{ssec:lag}}
To develop a variational model, we substitute in the
rescaled two-component Lagrangian
\begin{equation*}
\mathcal{L}=\int_{-\infty }^{\infty }\left( L_{1}+L_{2}+L_{12}+L_{21}\right)
dx,
\end{equation*}%
where
\begin{eqnarray}
L_{j} &=&E_{j}+\frac{i}{2}\left( u_{j}\frac{\partial \ u_{j}^{\ast }}{%
\partial t}-u_{j}^{\ast }\frac{\partial u_{j}}{\partial t}\right),  \notag
\\
E_{j} &=&\frac{1}{2}\left\vert \frac{\partial u_{j}}{\partial x}\right\vert
^{2}+V(x)\left\vert u_{j}\right\vert ^{2}+\frac{1}{2}\left\vert
u_{j}\right\vert ^{4},  \notag \\
L_{12} &=&L_{21}=\frac{1}{2}g\left\vert u_{1}\right\vert ^{2}\left\vert
u_{2}\right\vert ^{2},  \notag
\end{eqnarray}%
the Gaussian ansatz of the form
\begin{eqnarray}
u_1(x,t)&=&Ae^{-\frac{(x-B)^2}{2W^2}}e^{i(C+Dx+Ex^2)},\label{anz1}\\
u_2(x,t)&=&Ae^{-\frac{(x+B)^2}{2W^2}}e^{i(C-Dx+Ex^2)}.\label{anz2}
\end{eqnarray}
Starred variables indicate complex conjugation. The
parameters $A$, $B$, $C$, $D$, $E$, and $W$ are assumed to be time
dependent and they represent the amplitude, position, phase, wave
number, chirp, and width of the Gaussian ansatz, respectively. For
large $\Omega$, the steady state solution very closely resembles two
Gaussian functions separated by a distance $2B$, with constant
rotation of the phase $C=\mu t$, where $\mu$ is the chemical
potential (without loss of generality
we take $\mu=1$).
Upon interaction between the species and with the
harmonic trap, acceleration induces an inhomogeneity in the carrier
wave, known as chirp, that accordingly affects the phase
of each species.
It is important to note here that
the ansatz is invariant upon transposition of
space and atomic species component $x\rightarrow-x$ and
$u_1\rightarrow u_2$.
This allows us to reduce the number of parameters in the
system to six instead of having twelve, which would take into
account independent variation of the parameters in $u_1$ and $u_2$.
However, this simplification comes at a certain cost as, in particular,
it is not possible to monitor asymmetric (between the two components)
states within this ansatz; the latter type of states will be
partially explored within the dynamics of the species in
Sec.~\ref{sec:dynamics}.

When the Lagrangian is evaluated for the proposed ansatz, a
spatially-averaged effective Lagrangian is obtained
\begin{eqnarray}
{\cal L}&=&-{\sqrt{\pi}A^2}{W}\left[\Omega^2B^2+2E^2W^2+(D+2BE)^2
\phantom{\frac{dC}{dt}}\right.
\nonumber
\\
\displaystyle &&\displaystyle+\frac{A^2}{\sqrt{2}}(1+ge^{-\frac{2B^2}{W^2}})+\frac{1}{2W^2}+\frac{\Omega^2W^2}{2}
\nonumber
\\
\displaystyle
&&\displaystyle
\left.+2\frac{dC}{dt}+2B\frac{dD}{dt}+\frac{dE}{dt}(2B^2+W^2)\right],
\label{lag_avg}
\end{eqnarray}
and the equations of motion for the parameters are
obtained through the corresponding Euler-Lagrange equations
\begin{eqnarray}
\frac{\partial L}{\partial p_j}-\frac{d}{dt}\left(\frac{\partial L}{\partial \dot{p_j}}\right)=0,\label{ELeq}
\end{eqnarray}
%
%
where the parameter $p_{j}$, $j=1,2,\ldots,6$ represents the parameters in
the ansatz $A$, $B$, $C$, $D$, $E$, $W$ and $\dot{p_{j}}=dp_{j}/dt$.
The Euler-Lagrange equations for $A$, $B$, and $W$ evaluate to
$\partial L/\partial A=0$,
$\partial L/\partial B=0$ and
$\partial L/\partial W=0$
since the
second term in the Euler-Lagrange equation (\ref{ELeq})
is zero for these. This results in
equations that involve linear combinations of
$dC/dt$, $dD/dt$ and $dE/dt$
that when solved, give
Eqs.~(\ref{fiveode_C}), (\ref{fiveode_D}) and (\ref{fiveode_E}).
The rest of the
Euler-Lagrange equations give equations that involve linear combinations of
$dA/dt$, $dB/dt$ and $dW/dt$
that can
be solved to give
Eqs.~(\ref{fiveode_A}), (\ref{fiveode_B}) and (\ref{fiveode_W}).
The following equations are the result of solving the Euler-Lagrange
equations for the time derivatives of all the parameters of our ansatz:
\begin{eqnarray}
 \label{fiveode_A}
\frac{dA}{dt} &=& -A E,\\ \label{fiveode_B}
\frac{dB}{dt} &=& D+2BE,\\ \label{fiveode_C}
\frac{dC}{dt} &=& \frac{B^2}{2W^2}-\frac{D^2}{2} - \frac{1}{2W^2}+
\frac{\sqrt{2}A^2}{8W^2}(2B^2-5W^2)+
\nonumber\\&&\frac{\sqrt{2}A^2g}{8W^2}e^{-\frac{B^2}{2W^2}}(8B^4+2B^2W^2+5W^4),\\[1.0ex]
\label{fiveode_D}
\frac{dD}{dt} &=&
\frac{\sqrt{2}A^2Bg}{2W^4}e^{-\frac{B^2}{2W^2}}(4B^2+W^2) -\nonumber
\\&&
\frac{\sqrt{2}A^2b}{2W^2}- \frac{B}{W^4} - 2DE,\\[1.0ex]
\label{fiveode_E}
\frac{dE}{dt} &=&
\frac{\sqrt{2}A^2g}{4W^4}e^{-\frac{B^2}{2W^2}}(-4B^2+W^2)
+\nonumber\\&& \frac{\sqrt{2}A^2}{4W^2} + \frac{1}{2W^4} - 2E^2 -
\frac{\Omega^2}{2},\\ \label{fiveode_W}
\frac{dW}{dt} &=& 2EW.
\end{eqnarray}

\begin{figure}[htbp]
\includegraphics[width=8.0cm]{\rootfig 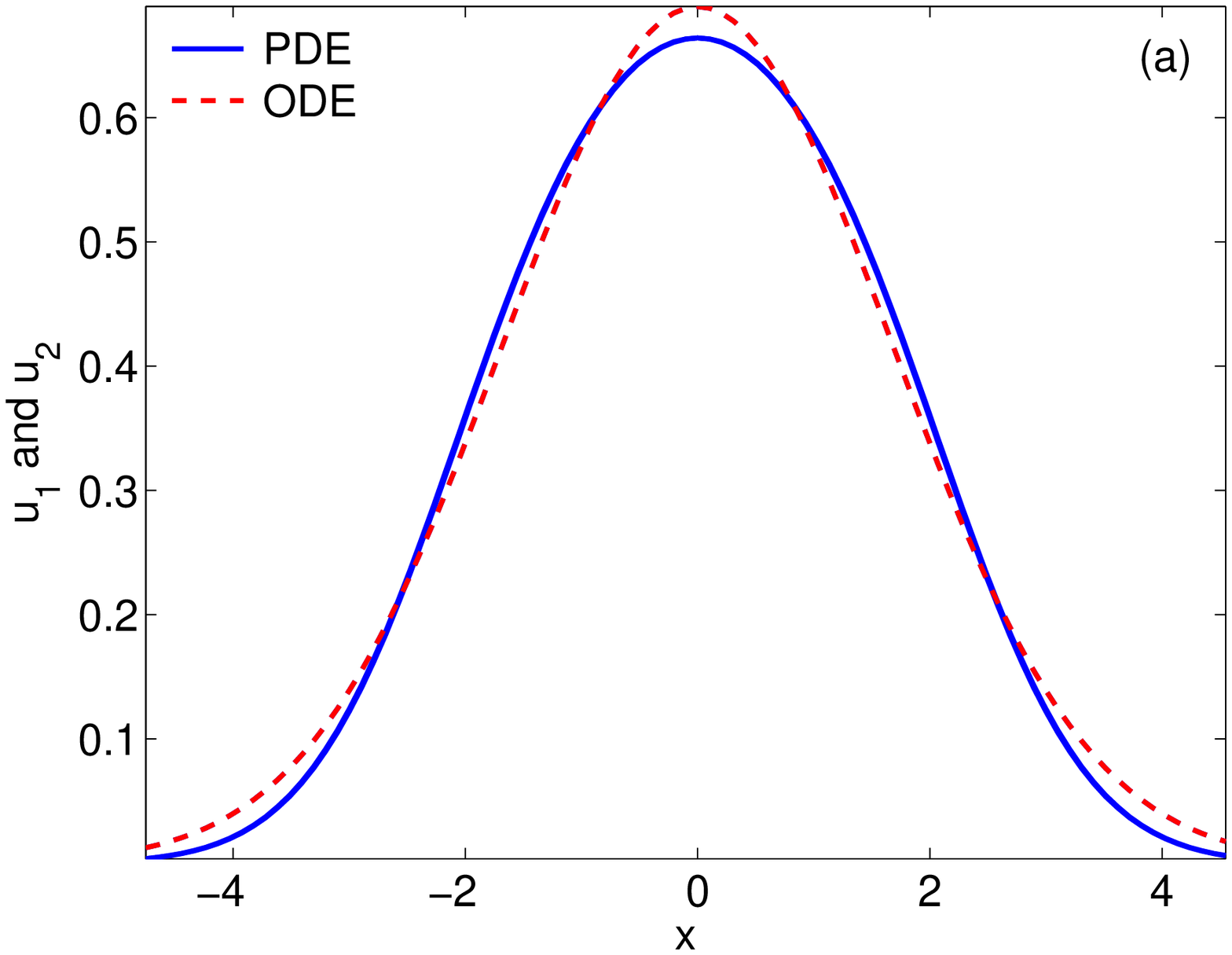}\\
\includegraphics[width=8.0cm]{\rootfig 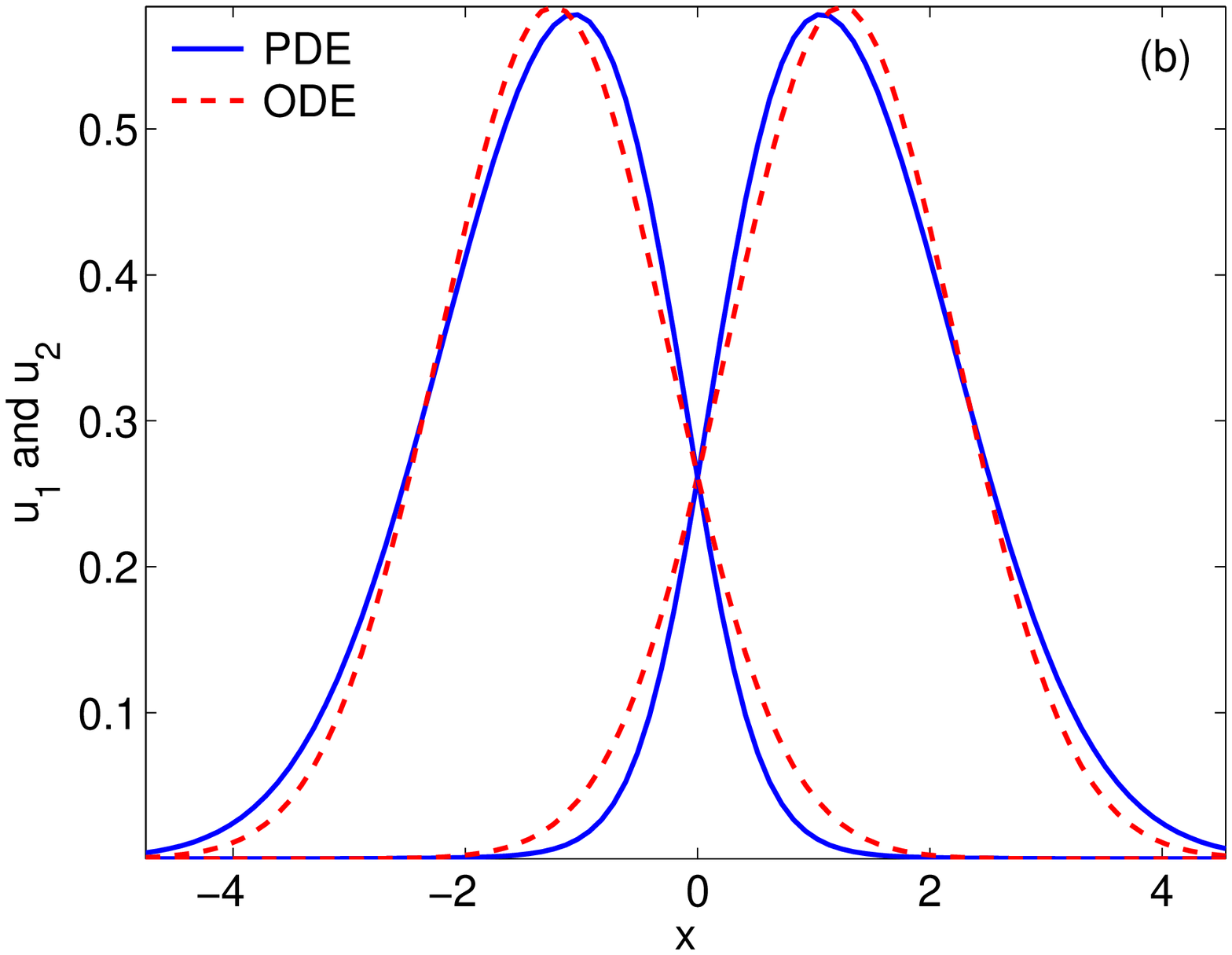}
\caption{(Color online)
(a) Steady state solution, $u_1$ and $u_2$,
for the mixed state when $B=0$.
(b) Steady state solution for the separated state when $B\neq0$.
The solid (blue) line
represents the steady state of the GP equations and the dashed (red)
line is the
steady state solution of the system of ODEs. Here, $\Omega=0.6$, $\mu=1$
$g=1$ for (a) and $g=20$ for (b).}\label{u1u2}
\end{figure}

As described below, these ODEs reflect fairly accurately the
principal dynamical features of the system. In particular, they capture the
oscillations of the two species  past each other when the
equilibrium separation $B$ is zero and the oscillations about their
corresponding phase-separated equilibrium position when
the two components are phase-separated.
\section{Steady State Solutions\label{sec:SSS}}
\subsection{Phase Bifurcations\label{ssec:bif}}
The steady state of Eqs.~(\ref{fiveode_A})--(\ref{fiveode_W})
is obtained by setting
$dA/dt=dB/dt=dC/dt=dE/dt=dD/dt=dE/dt=dW/dt=0$ witch leads immediately
to the
steady state solution $E_*=D_*=0$ and $C_*=\mu$. When the
equilibrium separation between the species is zero, the equilibrium
amplitude and width reduce to
\begin{eqnarray}
B_*&=&0,\label{bstar}\\
A_*^{2}&=&\frac{2\sqrt{2}\left(8\mu-\sqrt{15\Omega^2+4\mu^2}\right)}{15(1+g)},\label{astar}\\
W_*^{2}&=&\frac{\left(2\mu+\sqrt{15\Omega^2+4\mu^2}\right)}{5\Omega^2}.\label{wstar}
\end{eqnarray}
When the equilibrium separation is nonzero, the resulting
equilibrium amplitude and width are given by the transcendental
relations
\begin{eqnarray}
\Omega^2
-\frac{\sqrt{2}A_*^2g}{W_*^2}e^{-\frac{B_*^2}{2W_*^2}}&=&0,\label{steady1}\\
\mu-\frac{1}{2W_*^2}-\frac{5W_*^2}{8}\left(\sqrt{2}A_*^2g+\Omega^2W_*^2\right)&=&0,\label{steady2}\\
\mu+\frac{3}{4W_*^2}-\frac{5}{4}\Omega^2\left(W_*^2+2B_*^2\right)&=&0.\label{steady3}
\end{eqnarray}

Figure \ref{u1u2} shows that the steady state solution of the
full GP model closely
matches the steady state from the ODEs, indicating that the ansatz
successfully captures the relevant PDE behavior. For large values of $\Omega$,
the steady state solution of the GP deviates from the Gaussian
shape, resembling an inverted parabola, often referred to as the
Thomas-Fermi approximation \cite{Pethick-book,stringari}.

\begin{figure}[htbp]
\begin{center}
\includegraphics[width=8cm,height=5.5cm]{\rootfig 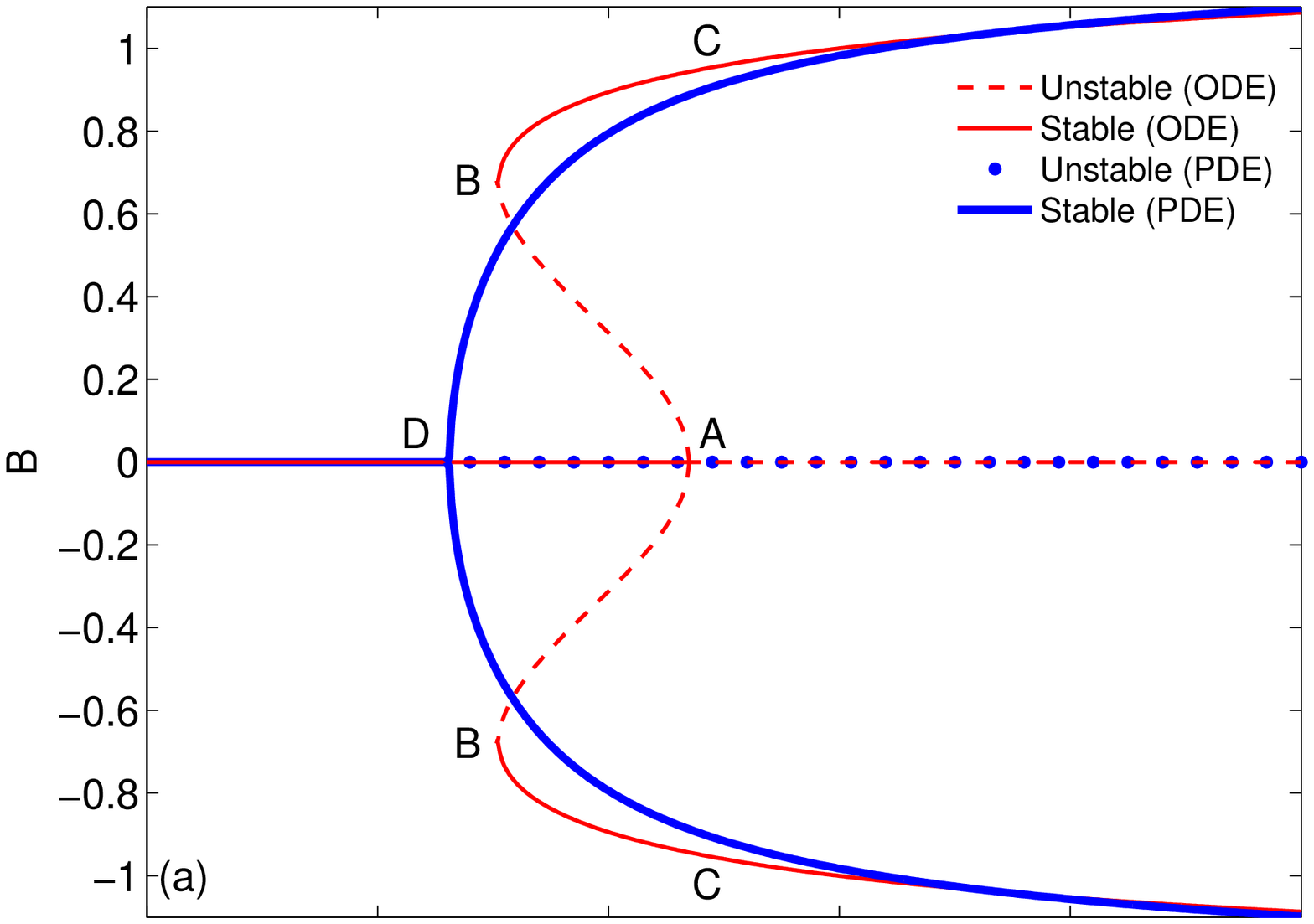}\\
\includegraphics[width=8cm,height=5.5cm]{\rootfig 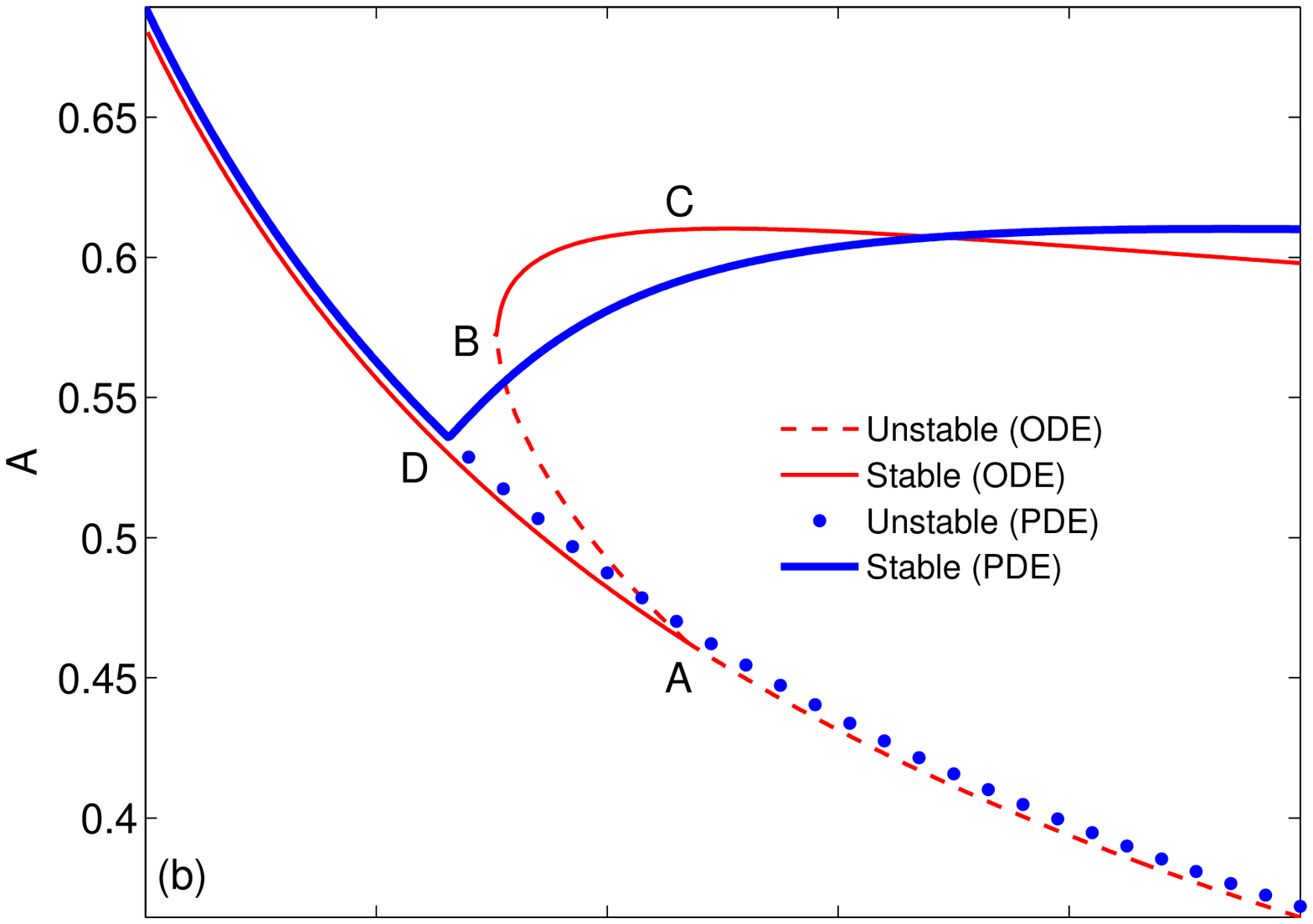}\\
\includegraphics[width=8cm,height=5.5cm]{\rootfig 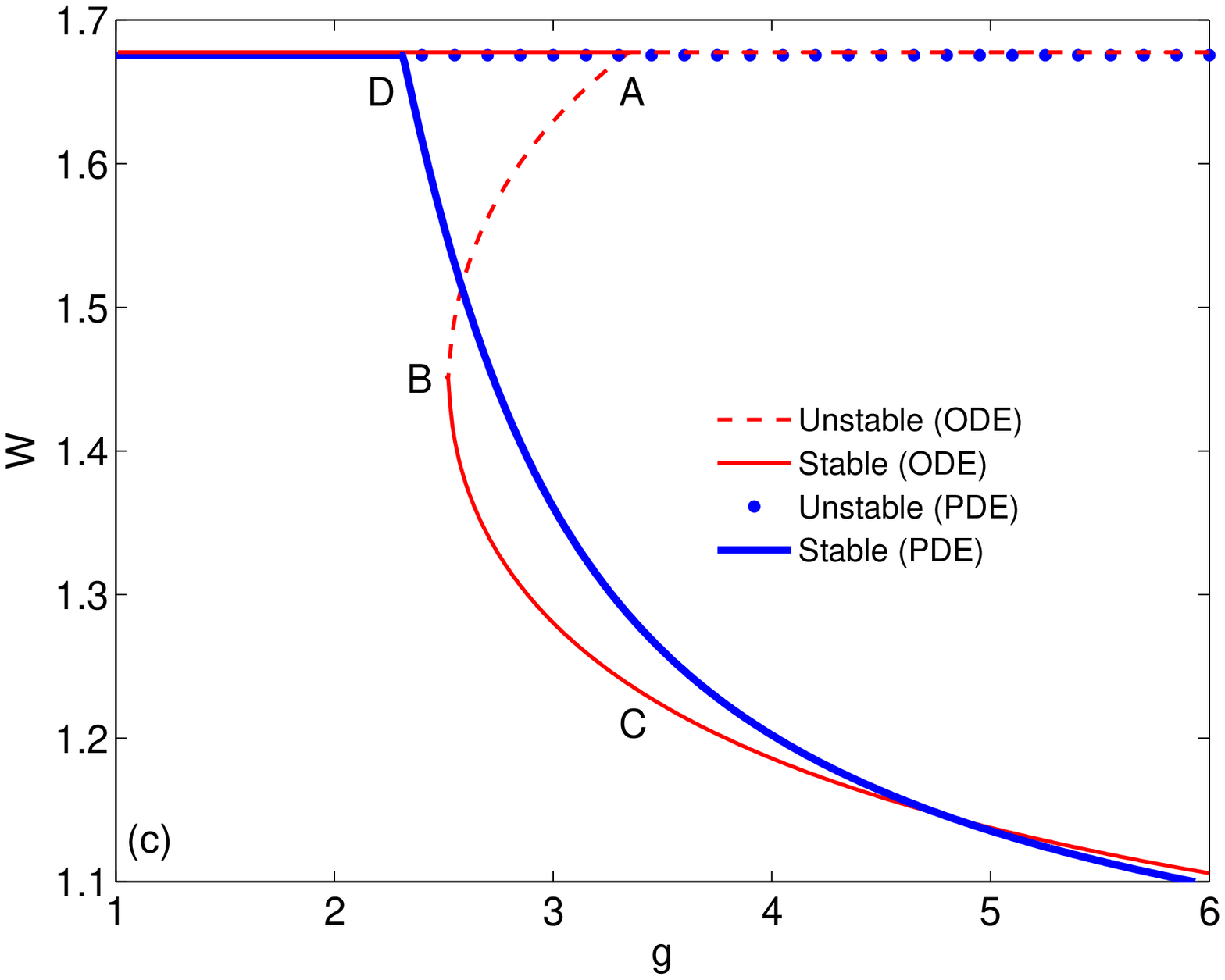}
\end{center}
\vspace{-0.4cm}
\caption{(Color online)
Equilibrium (a) separation, (b) amplitude, and
(c) width for the two condensed species, as a function
of the interspecies coupling strength $g$ for
$\Omega=0.6$ and $\mu=1$. The (red) thin solid (stable)
and dashed (unstable) lines correspond to the steady states
for the reduced ODE model [Eqs.~(\ref{fiveode_A})-(\ref{fiveode_W})],
while the (blue) thick solid (stable) and dots (unstable) correspond
to the steady state from the full GP model
[Eqs.~(\ref{GPE1_1d})-(\ref{GPE2_1d})].
}\label{bifabw}
\end{figure}

Detailed bifurcation diagrams for the amplitude, position, and width
of each species can be obtained by solving
Eqs.~(\ref{steady1})--(\ref{steady3}) for $A$, $B$, and $W$ for the phase
separated state and Eqs.~(\ref{bstar})--(\ref{wstar}) for the mixed
state.
The steady state solution reveals a pitchfork bifurcation
for the position of each condensate as the interspecies coupling
strength is increased as can be seen in Fig.~\ref{bifabw}a.
Interestingly, the
steady state of the GP equations produces a {\it supercritical} pitchfork
bifurcation at point D of Fig.~\ref{bifabw}. More specifically,
for small values
of $g$, a stable mixed phase can be identified; as $g$ is increased, the
mixed phase becomes unstable and the phase separated state becomes
stable.
On the other hand, the system of ODEs
also predicts a pitchfork bifurcation,
however the approximate nature of the ansatz results in the
identification of the bifurcation as a {\it subcritical}
one (point A) occurring in the vicinity of a symmetric pair of saddle
node bifurcations (points B).
Both bifurcation diagrams agree with each other away
from the transition between phases. In the vicinity of the
transition point, clearly, the nature of the ansatz is
insufficient to capture the fine details of the bifurcation
structure (thus inaccurately suggesting a phenomenology
involving bistability, and hysteretic behavior of the system).
It should be noted that
the phase separation from the GP model (point D) lies near the
saddle node bifurcation point (point B) and the subcritical
bifurcation point (point A) from the system of ODEs. At small values
of the harmonic trap strength, the true point of phase transition is
closer to the subcritical bifurcation point and at larger values
of the harmonic trap strength, the true point of phase transition
lies closer to the saddle node bifurcation point.

\subsection{Phase Separation\label{ssec:phas}}
\begin{figure}[htbp]
\hskip0.15cm  \includegraphics[width=8.5cm]{\rootfig 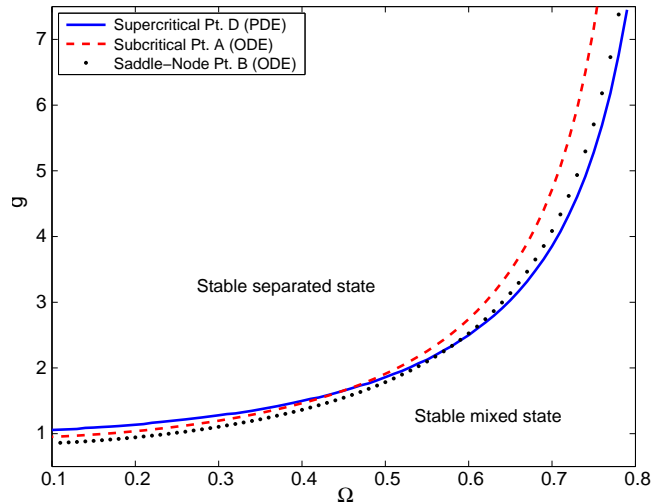}
\vspace{-0.3cm}
\caption{(Color online)
The (blue) solid line represents the boundary of
zero species separation from the GP equations
(point D in Fig.~\ref{bifabw}),
where states that
lie to the right of this line are mixed and values to the left are
phase-separated. The (red) dashed line represents the boundary of zero
species separation for the system of ODEs given by
Eq.~(\ref{phase_separ}) (point A in Fig.~\ref{bifabw}).
The (black)
dotted line shows the saddle-node point (point B in Fig.~\ref{bifabw})
obtained from the system of ODEs.}\label{phase_diagram1}
\end{figure}

Using the zero separation amplitude and width, point A in
Fig.~\ref{bifabw}, an expression
for the onset of phase separation can be approximated in terms of the
critical interspecies interaction
\begin{equation}
g_{\rm cr}=
\frac{6\mu+3\sqrt{15\Omega^2+4\mu^2}}{26\mu-7\sqrt{15\Omega^2+4\mu^2}}.\label{phase_separ}
\end{equation}
Despite the deviation of point A in Fig.~\ref{bifabw}, from the relevant
point D of the corresponding PDE, the analytical expression offers
valuable insight on the dependence of the critical interspecies
interaction for phase separation on parameters of the trap (in particular,
its frequency) and those of the condensate (in particular, its chemical
potential); see also Fig.~\ref{phase_diagram1} for a detailed
comparison of the ODE and PDE bifurcation points. More specifically,
the equation predicts that when the harmonic trap's frequency
approaches zero, $\Omega\rightarrow0$, phase separation occurs when
$g_{\rm cr}\rightarrow 1$ coinciding with the
well known miscibility condition $\Delta=g^2-1=0$.
As $\Omega\rightarrow \Omega_{\rm cr}=4 \sqrt{2}\mu/7 \approx 0.8\mu$ phase
separation occurs at $g_{\rm cr}\rightarrow \infty$.
This behavior can be qualitatively understood since
tighter (larger $\Omega$) traps tend to ``squeeze'' both
components together, thus frustrating the system's
tendency towards phase separation.

\begin{figure}[htbp]
\begin{center}
\includegraphics[width=8.5cm]{\rootfig 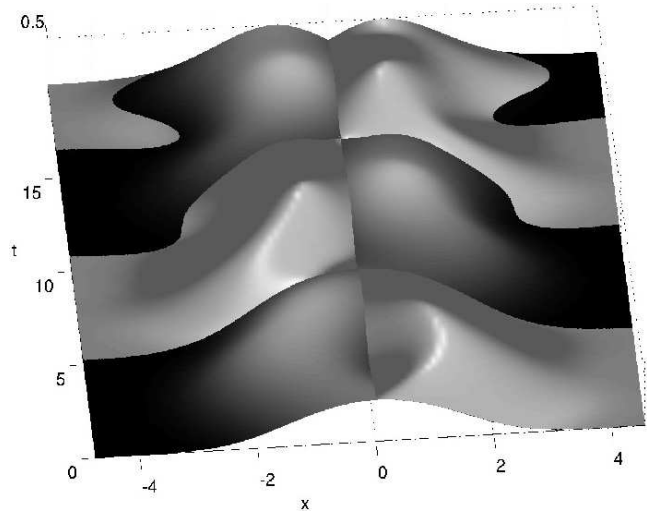}
\end{center}
\vspace{-0.4cm}
\caption{Oscillations for a mixed state from direct
numerical integration of the GP model (\ref{GPE1_1d})-(\ref{GPE2_1d}) for
$g=1$ ($g < g_{\rm cr}$), $\mu=1$ and $\Omega=0.6$. Lighter gray corresponds to one
the density of one component and darker gray to the other component.
}\label{time_it0A}
\end{figure}

\begin{figure}[htbp]
\begin{center}
\includegraphics[width=8.5cm,height=1.6cm]{\rootfig 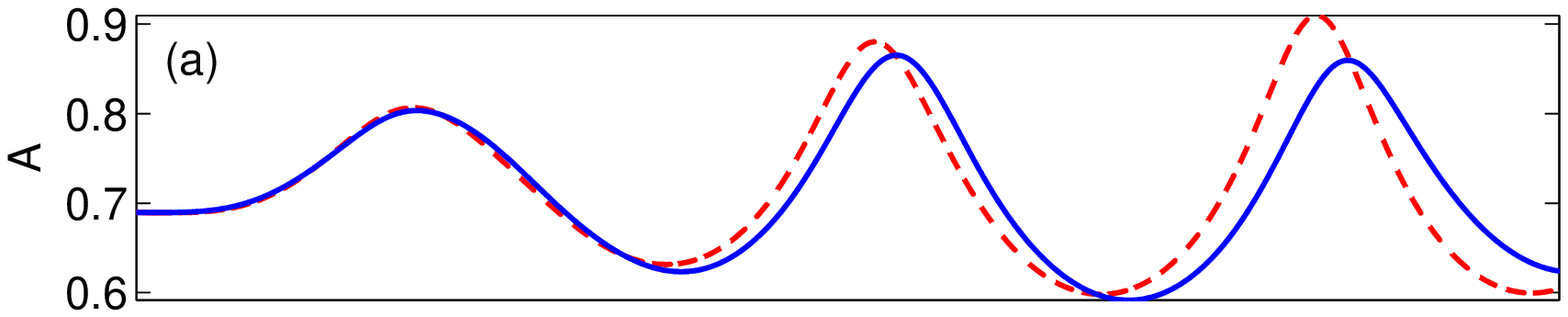}\\
\includegraphics[width=8.5cm,height=1.6cm]{\rootfig 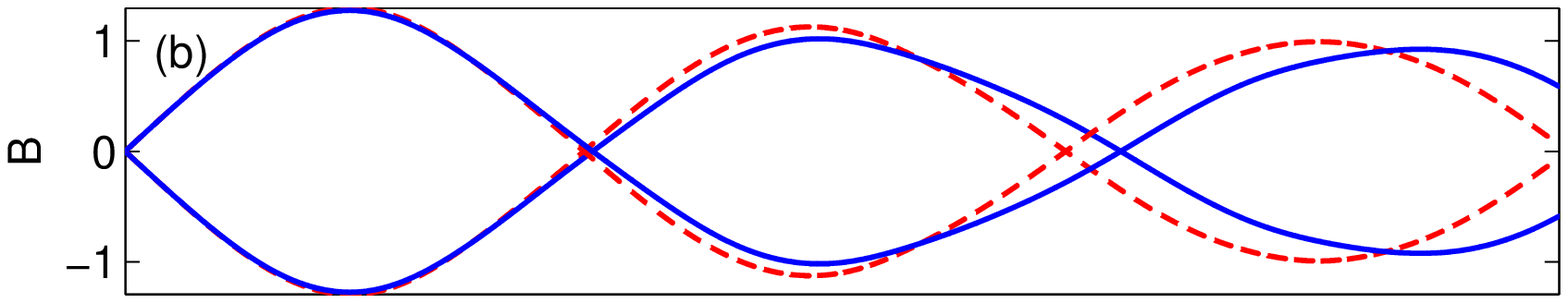}\\
\includegraphics[width=8.5cm,height=1.6cm]{\rootfig 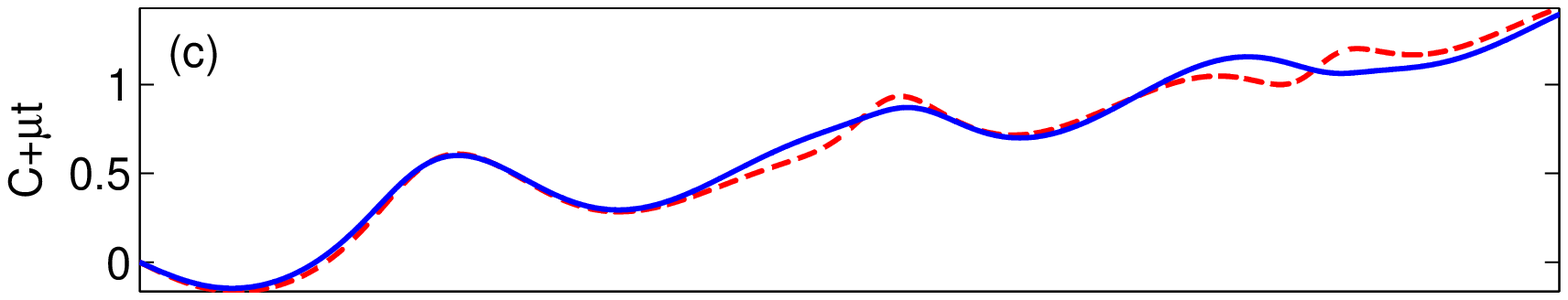}\\
\includegraphics[width=8.5cm,height=1.6cm]{\rootfig 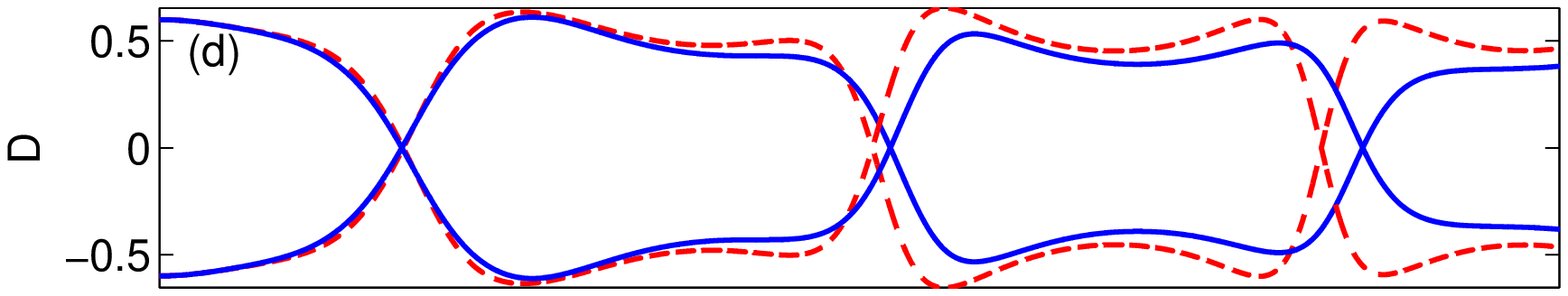}\\
\includegraphics[width=8.5cm,height=1.6cm]{\rootfig 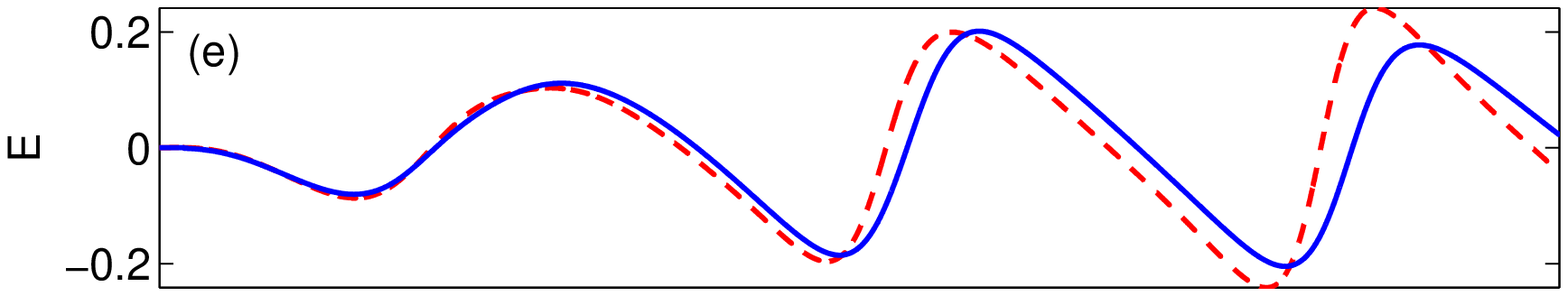}\\
\includegraphics[width=8.5cm,height=2.2cm]{\rootfig 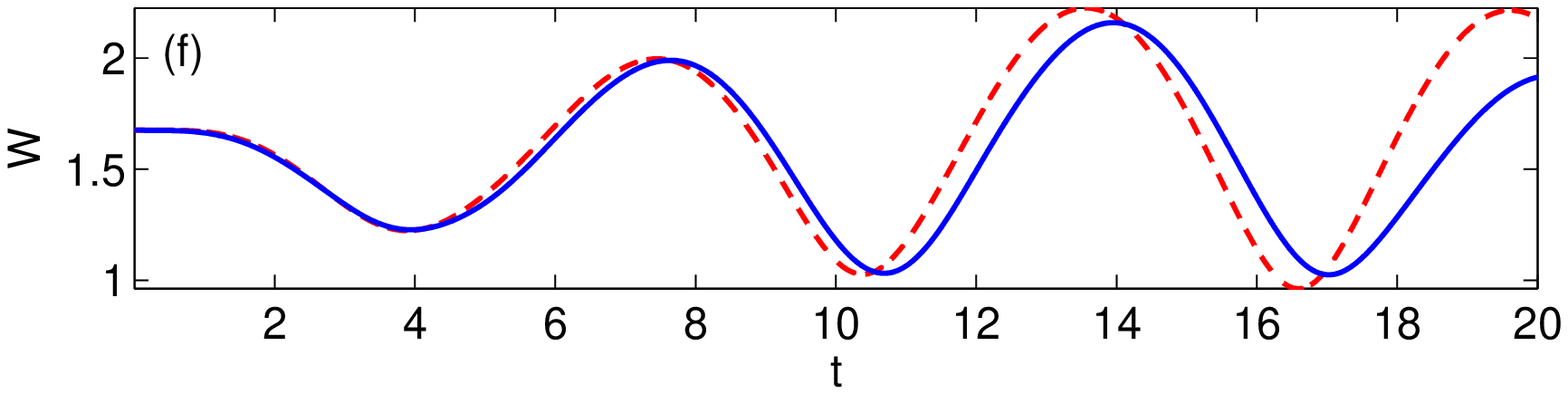}
\end{center}
\vspace{-0.4cm}
\caption{(Color online)
(a) Amplitude, (b) position, (c) phase, (d) velocity, (e) chirp,
and (f) width corresponding to the {\em mixed} oscillating
state show in Fig.~\ref{time_it0A}. Solid lines represent results
from direct numerical simulations of the GP equation while dashed
lines depict the results for the ODE reduction
(\ref{fiveode_B})--(\ref{fiveode_W}).
}\label{time_it0}
\end{figure}

\begin{figure}[htbp]
\begin{center}
\includegraphics[width=8.5cm]{\rootfig 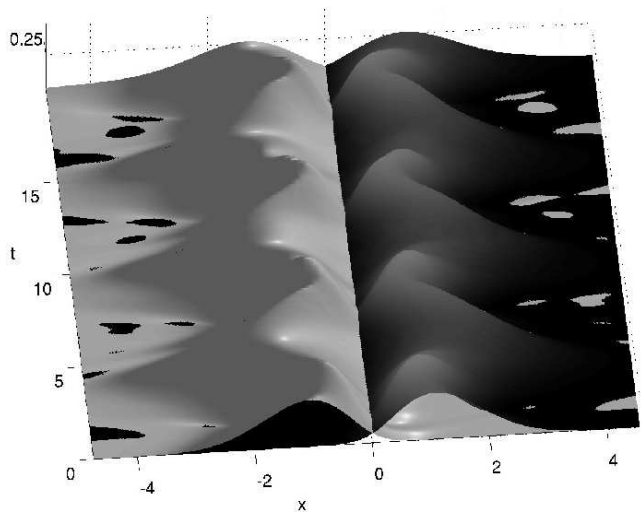}
\end{center}
\vspace{-0.4cm}
\caption{Same as in Fig.~\ref{time_it0A} for the
time evolution of oscillating species about their
phase separated configuration when $g=20$ ($g > g_{\rm cr}$).
}\label{time_it2A}
\end{figure}

\begin{figure}[htbp]
\begin{center}
\includegraphics[width=8.5cm,height=1.6cm]{\rootfig 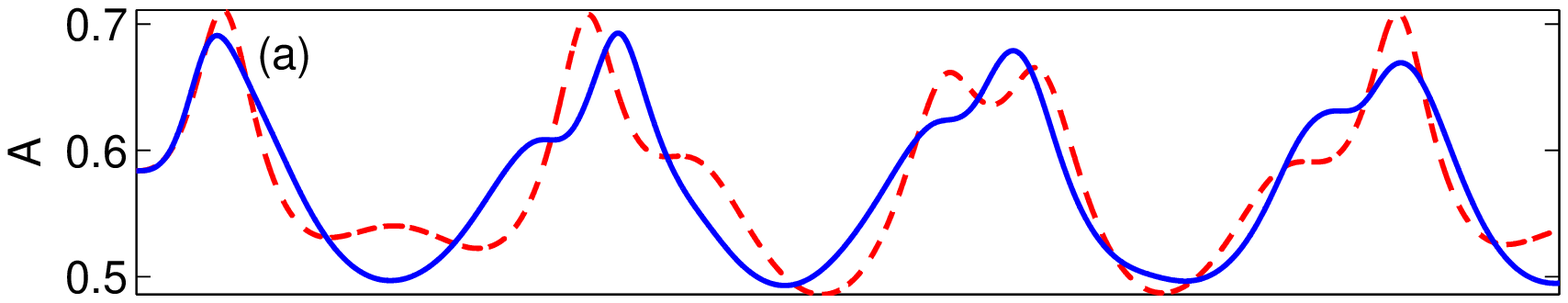}\\
\includegraphics[width=8.5cm,height=1.6cm]{\rootfig 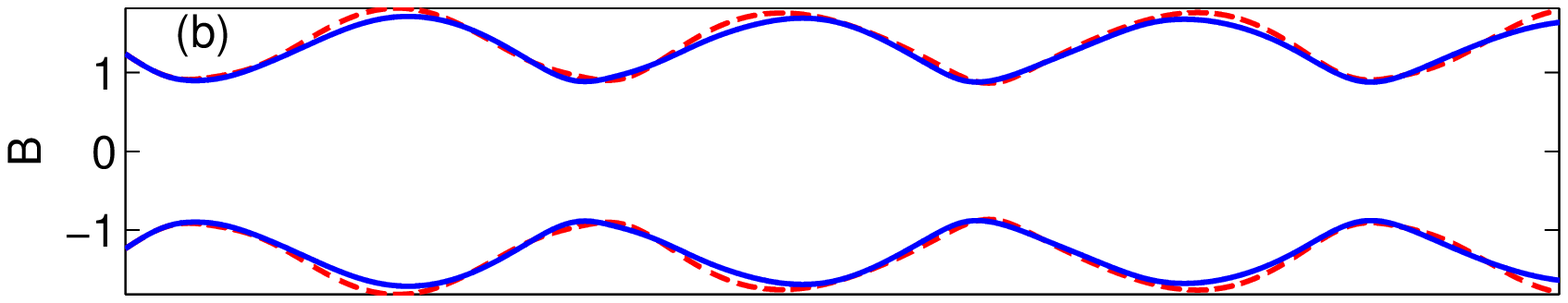}\\
\includegraphics[width=8.5cm,height=1.6cm]{\rootfig 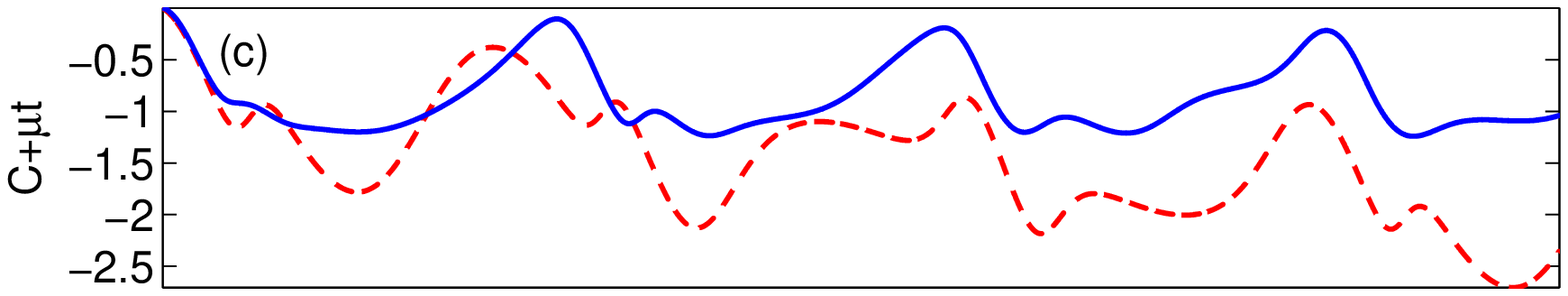}\\
\includegraphics[width=8.5cm,height=1.6cm]{\rootfig 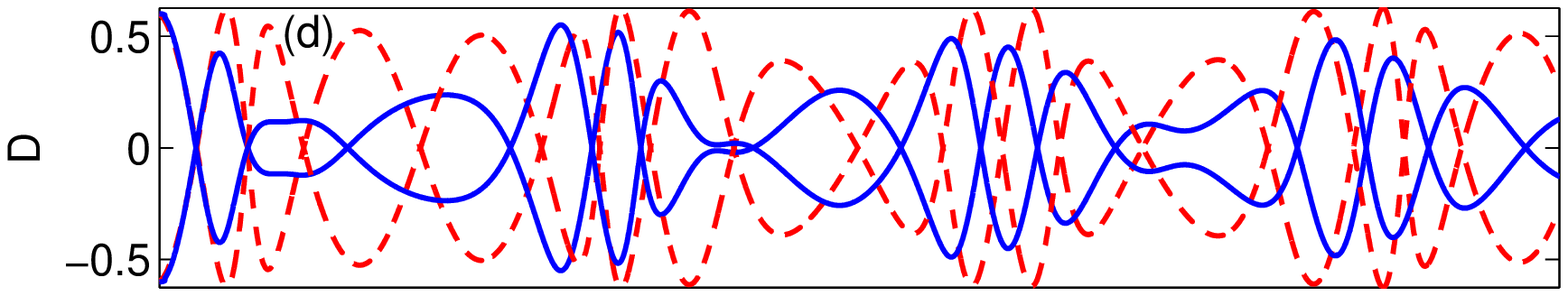}\\
\includegraphics[width=8.5cm,height=1.6cm]{\rootfig 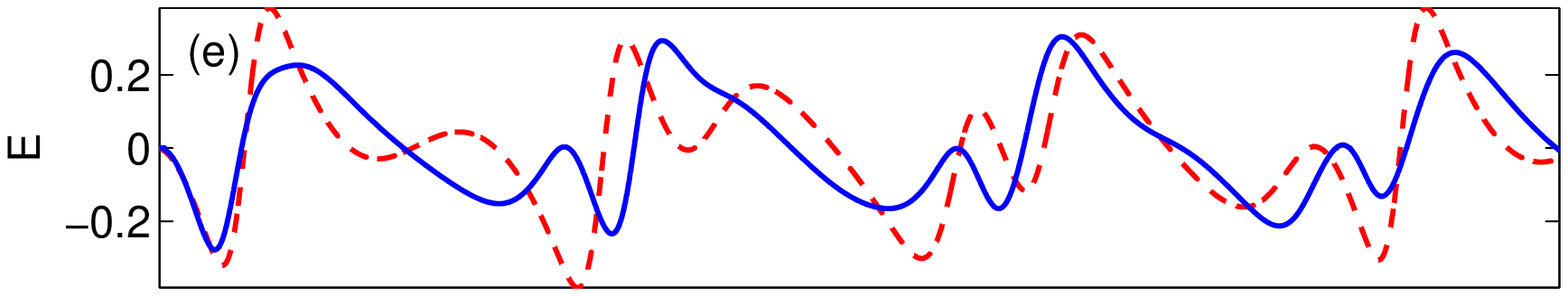}\\
\includegraphics[width=8.5cm,height=2.2cm]{\rootfig 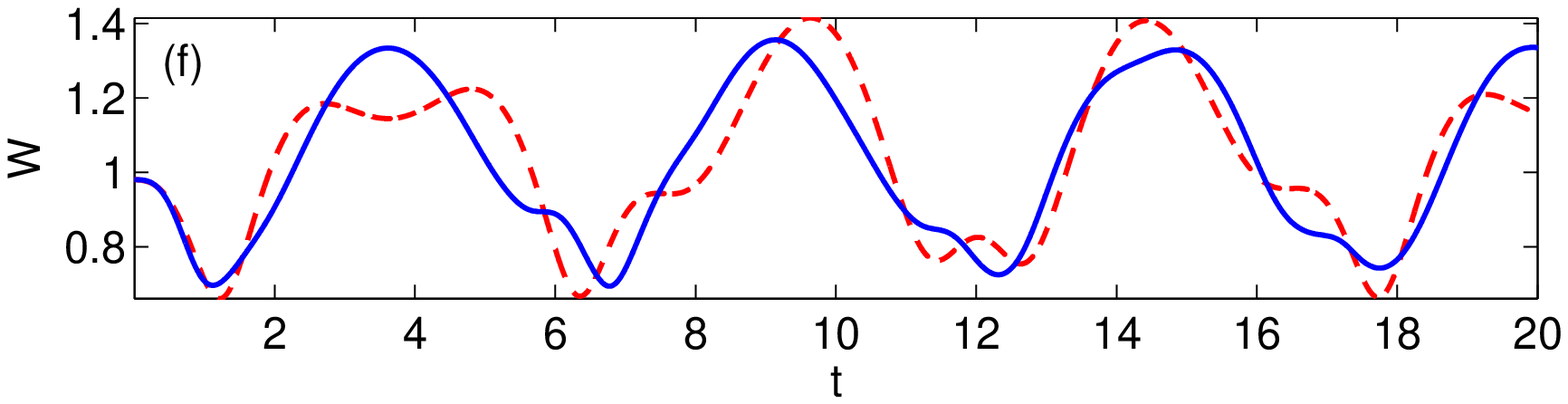}
\end{center}
\vspace{-0.4cm}
\caption{(Color online)
Same as in Fig.~\ref{time_it0} for the
time evolution of oscillating species about their
phase separated configuration as shown in Fig.~\ref{time_it2A}.
}\label{time_it2}
\end{figure}
The prediction of the system of ODEs for the location of
the bifurcation point
agrees well with the results from the GP model
for small values of the harmonic trap strength. Recall that
the phase separation
for the ODE model is located at the subcritical pitchfork
bifurcation point A in Fig.~\ref{bifabw}.
However, as can be observed from Fig.~\ref{phase_diagram1},
a better approximation for the full system's phase separation
(point D), in the case of large trapping frequencies,
can be given by the saddle-node bifurcation
point B for the ODE reduced system for $\Omega$ values
close to $\Omega_{\rm cr}$.

\subsection{Dynamics of the Reduced System\label{ssec:dyn}}
For relatively small values of the interspecies coupling $g$
(i.e., to the left of point D in Fig.~\ref{bifabw}) the two species
do not separate and thus, when given opposite direction
velocities from the mixed state, they will oscillate through each other
as depicted in Fig.~\ref{time_it0A}.
This case is analyzed in Fig.~\ref{time_it0} where
the two components oscillate through
each other about their common equilibrium separation of zero and the
prediction from the system of ODEs (red dashed lines)
agrees very well with the direct
integration of the GP equations (blue solid lines).
Because of conservation of mass, the
amplitude and width oscillate out of phase: the amplitude is
maximized and width is minimized when the acceleration of the
species is maximized; on the other hand, the width is
maximized and the amplitude is minimized when the velocity is maximized.
The velocity of each species [see Eq.~(\ref{fiveode_B})]
has two components, one that depends on the wave number $D$
and another that depends on the product of chirp and position $2EB$.
If there is no chirp, the wave number is the velocity of the
condensate species.  In Fig.~\ref{time_it0}c and \ref{time_it2}c, a
factor of $\mu t$ has been added to show the deviation of the phase from
the steady state value.

\begin{figure}[htbp]
\begin{center}
\hskip0.15cm \includegraphics[width=8.3cm]{\rootfig 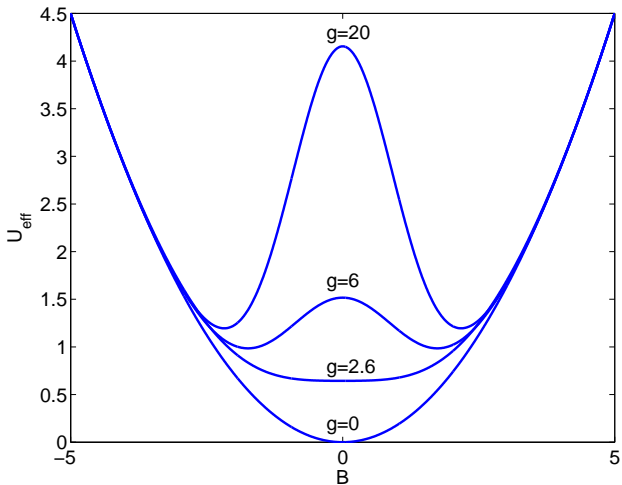}
\end{center}
\vspace{-0.5cm} \caption{(Color online)
The effective potential, $U_{\rm eff}$, for several values of the
species coupling parameter: $g=0$, $g=2.6$, $g=6$, and $g=20$. Here,
$\Omega=0.6$ and $\mu=1$.
}\label{pot4}
\end{figure}

\begin{figure}[htbp]
\begin{center}
\includegraphics[width=8.0cm]{\rootfig 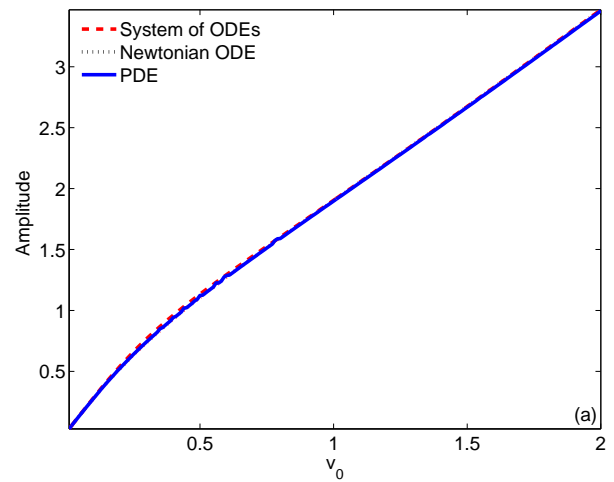}
\includegraphics[width=8.0cm]{\rootfig 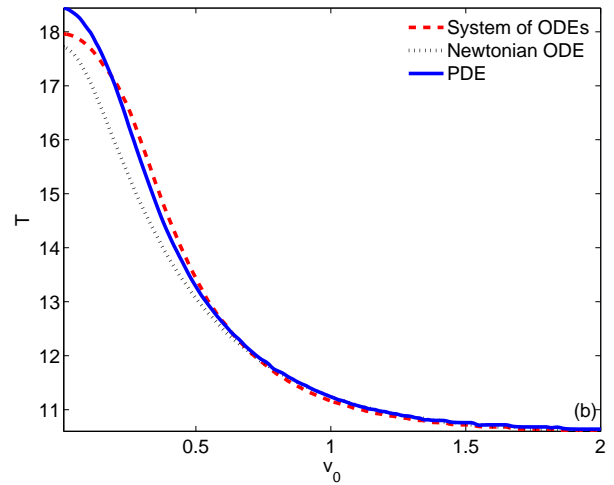}
\end{center}
\vspace{-0.5cm} \caption{(Color online)
(a) Amplitude of oscillation and (b) period of oscillation,
$T$, as a function of increasing initial velocity, $v_0$. The (blue) solid
line represents results from the coupled GP equations, the (red) dashed
line represents results from the system of ODEs, and the black dotted line
represents the results from the Newtonian reduction. Here, $g=1$,
$\mu=1$ and $\Omega=0.6$.
}\label{per}
\end{figure}

On the other hand, if we assume a well-separated state as
the PDEs initial condition,
(right of point D in Fig.~\ref{bifabw}), it is possible
that the two components entertain oscillations about their
phase-separated
steady state as depicted in Fig.~\ref{time_it2A}.
As illustrated in Fig.~\ref{time_it2}, the two phase separated components
collide against each other as they oscillate (but do not
go through each other) about a nonzero
position. The prediction from the system of ordinary differential
equations once again agrees very well with the
numerical integration of the GP
equation
for the position of the two species, but differs somewhat for the
other parameters of motion.  This occurs because the time evolution
of the solution of Eqs.~(\ref{GPE1_1d}) and (\ref{GPE2_1d})
in this case, due to the oscillation and interaction, deviates
from a Gaussian waveform and the corresponding variational
prediction begins to lose accuracy.

\begin{figure}[htbp]
\begin{center}
\includegraphics[width=8.2cm]{\rootfig 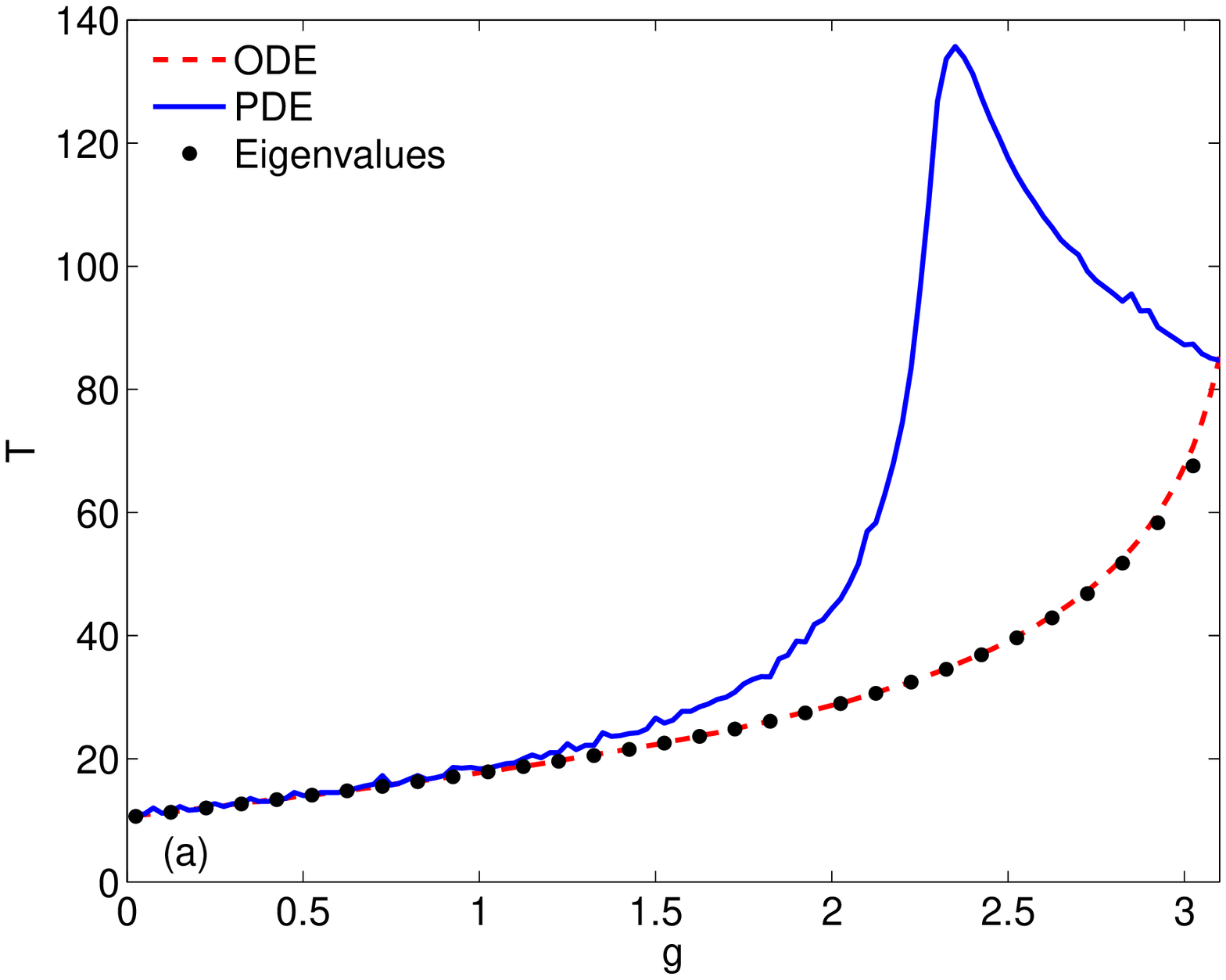}\\
~~\includegraphics[width=8.0cm]{\rootfig 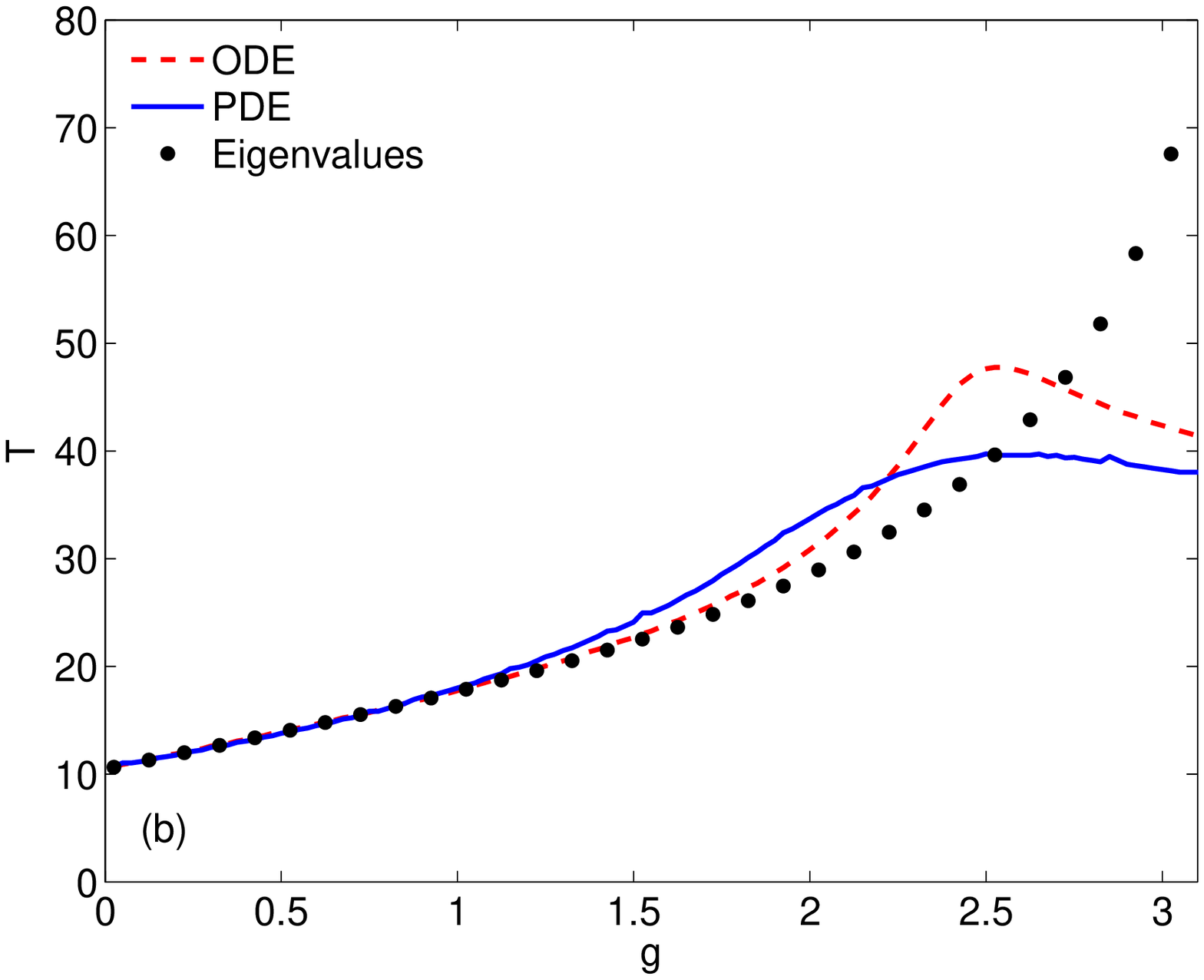}
\end{center}
\vspace{-0.5cm} \caption{(Color online)
(a) The period of oscillation, $T$, is shown as a
function of $g$ for an
initial velocity of $v_0=0.01$. (b) The period of oscillation, as a
function of $g$ for an initial velocity of $v_0=0.1$. The (blue) solid
line represents results from the coupled GP equations, the (red) dashed
line represents results from system of ODEs, while the black dots
represent the results from the evaluation of the eigenvalues of the
Jacobian of the system of ODEs at equilibrium. Here, $\Omega=0.6$
and $\mu=1$.
}\label{eig_osc}
\end{figure}

\subsection{Newtonian Reduction\label{ssec:newt}}
To develop a more tractable model for the dynamics, a
classical Newtonian system for the motion of the center of each
species is desirable. Taking a time derivative of Eq.~(\ref{fiveode_B})
and substitution of Eqs.~(\ref{fiveode_B}), (\ref{fiveode_D}),
and (\ref{fiveode_E}) yields
\begin{equation}
\frac{d^2B}{dt^2} = -\left(\Omega^2 -
\frac{\sqrt{2}A_*^2g}{W_*^2}e^{-\frac{B^2}{2W_*^2}}\right)B,
\label{newton_ode}
\end{equation}
where we simplified the dynamics by replacing the time
dependent variations of $A$ and $W$ by their respective
equilibrium values $A_*$ and $W_*$.
This simplification is justified by the fact that the
oscillations in $A$ and $W$ are relatively weak as
it can be observed from Figs.~\ref{time_it0}a,
\ref{time_it0}f, \ref{time_it2}a, and \ref{time_it2}f.
These phase separated oscillations contain two fundamental
physical features: the external
trapping potential and an exponential repulsive interaction that
depends on the cross species coupling parameter $g$. Integrating
Eq.~(\ref{newton_ode}) with respect to $B$ yields a Newtonian
equation of motion under the effective potential
\begin{equation}
U_{\rm eff}= \frac{\Omega^2}{2}B^2 +
\frac{\sqrt{2}A_*^2g}{2}e^{-\frac{B^2}{2W_*^2}}. \label{pot_eff}
\end{equation}
This reduced dynamics gives an effective double well potential for
a fully phase separated state ($g>g_{\rm cr}$) and a nonlinear single well
potential for the mixed states ($g<g_{\rm cr}$).
It is important to note that
in the expression of Eq.~(\ref{pot_eff}), $A_*$ and $B_*$
vary as a function of $g$.
%
This dependence has been incorporated in Fig.~\ref{pot4}
which shows the effective potential for increasing
values of $g$, where it transitions from a single well potential to
a double well potential. For $\Omega=0.6$ and $\mu=1$, the reduced
Newtonian model predicts that at $g=2.6$ the two species phase separate
and the potential becomes very flat yielding  very long periods of
oscillation. It is remarkable to point out that a similar picture
has been drawn for the interaction of two particle-like excited
states (i.e., dark solitary matter waves) within the same species and
has been found to be extremely successful in comparisons with experimental
results \cite{oberthaler}. In that case, as well, the fundamental
characteristics were the parabolic trapping and the exponential
tail-tail interaction between the waves.

\begin{figure*}[htbp]
\includegraphics[width=16cm,height=10cm]{\rootfig 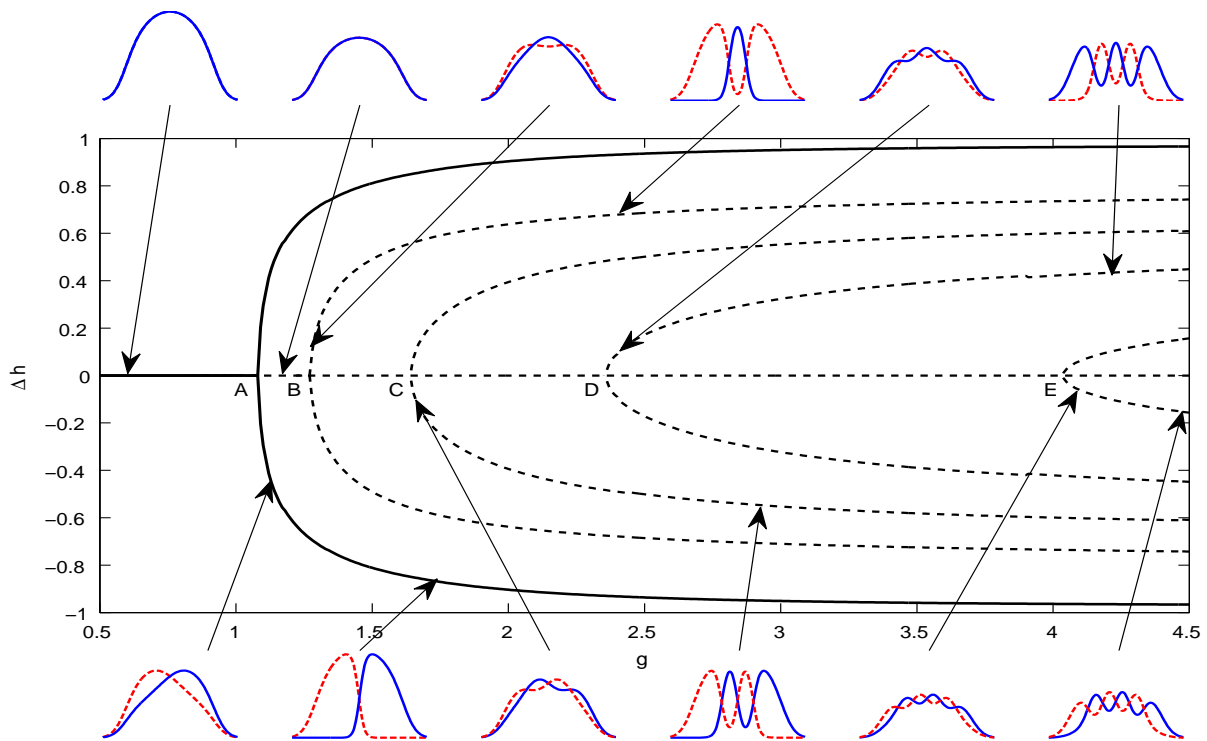}
\vspace{-0.2cm}
\caption{(Color online)
Degree of phase separation $\Delta h=u_1(x_c)-u_2(x_c)$,
where $x_c$ is the location of the closest density
maximum (irrespective of component) to the trap center,
as a function of
the interspecies coupling parameter for various excited states
for $g \in [0.5,4.5]$ and $\Omega=0.2$. Stable (unstable) solutions
branches are depicted with black solid (dashed) lines.
Typical solutions for each branch are depicted in the
surrounding insets.}
\label{exbif}
\end{figure*}

Figure \ref{per} depicts both the (a) amplitude and (b)
period of oscillations predicted by the Newtonian reduction
in comparison with the corresponding PDE findings.
It is seen that the GP equation and the ODE system agree
very well for a wide range of values of initial velocities. This figure
also shows the nonlinearity of oscillations where small amplitude
oscillations have a period of 18.3 and large oscillations yield a
period of $T\approx 10.5$, which corresponds to
the harmonic trap's period $T\rightarrow {2\pi}/{\Omega}$.
Figure \ref{eig_osc}a shows that the period of oscillation
from the system of ODEs matches that of the GP model for small values of
$g$.  As $g\rightarrow 2.4$ (i.e., when approaching phase separation
of the GP equations [point D in Fig.~\ref{bifabw}]),
the period changes substantially. Then as $g\rightarrow
3.1$, the system of ODEs begins to phase-separate.
Since these are small oscillations, the eigenvalues from the
Jacobian of the system of ODEs at equilibrium matches very well the
oscillations of the system of ODEs.
Figure \ref{eig_osc}b shows similar results to
Fig.~\ref{eig_osc}a  but for larger oscillations.
We can see that for larger oscillations the
ODEs' period more closely matches that of the GP
for smaller $g$, while for larger values of the interspecies
strength, the deviation becomes more significant.
Furthermore, for larger values of $g$, the period obtained from the
eigenvalues also deviates from the period from the ODEs. For large
oscillations, the two species do not effectively interact
(since they go through each other too rapidly to
``feel'' each other) and the period is roughly independent
of $g$ as predicted by Eq.~(\ref{newton_ode}) when $|B|\ll1$,
yielding simple harmonic
oscillations. In this case, the GP's period is close to that predicted by the
ODEs.

\begin{figure}[htbp]
\begin{center}
\includegraphics[width=8.5cm]{\rootfig 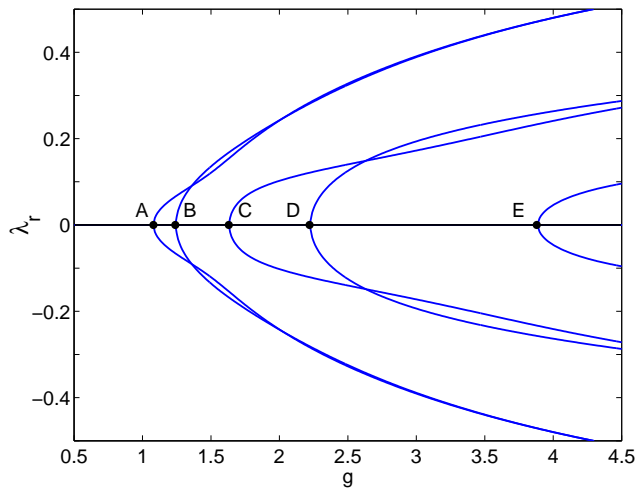}
\vspace{-0.2cm}
\caption{(Color online)
Real part of largest eigenvalue of the
mixed state as a function of the interspecies coupling parameter for
the various excited states depicted in Fig.~\ref{exbif}
with $g \in [0.5,4.5]$, and $\Omega=0.2$.}\label{exphase}
\end{center}
\end{figure}

\section{Excited States\label{sec:ES}}
%
%
For sufficiently weak traps
$\Omega<0.5$, as the interspecies coupling strength is
increased for
$g>1$, new, high order mixed states emerge. These higher excited
states correspond to alternating bands dominated
successively by each of two  species.
As $\Omega$
is decreased or alternatively $g$ is increased, the number of alternating
bands increases within the solution profile
and the population imbalance within each band less pronounced.
%
Each solution branch is found by using parameter continuation
on the parameter $g$ using a Newton fixed point iteration to
find the stationary state.
The stability for each computed profile was obtained
by computing the eigenvalues of the linearized dynamics
at the fixed points using standard techniques \cite{Our-book}.
We now describe
the series of bifurcations that occur as
the interspecies coupling $g$ increases as depicted in Fig.~\ref{exbif}:
\begin{itemize}
\item
For $g<1.08$, the only
state that exists is the mixed state and it is stable.
This threshold is close to the traditional miscibility
condition $\Delta=g^2-1=0$.
\item
At $g=1.08$ (point A),
the mixed states becomes unstable, through the previously described
supercritical pitchfork bifurcation, leading to the emergence of
a stable phase-separated state where one component moves to the
left and the other one moves to the right.
%
\item
At $g=1.24$ (point B), a second supercritical pitchfork occurs,
rendering the symmetric (across components) solution more
unstable and giving rise
to a state where one species has a single hump
and the other one has a double hump. We call this state a
1-2 hump configuration.
\item
At $g=1.63$ (point C), the third supercritical pitchfork bifurcation
of the series arises, leading this time to a situation where
one species moves to the left and the other one moves to the right, both
forming a double hump (a 2-2 hump configuration).
\item
At $g=2.22$ (point D), a double hump with a triple hump state arises
(a 2-3 hump configuration).
\item
At $g=3.8$ (point E), a triple hump with a triple hump state forms
(a 3-3 hump configuration).
\end{itemize}
As the
condensate becomes wider, more and more bands appear, each band
having a width comparable to the healing length of the condensate.
This picture seems to be natural from the point of view of \cite{tsubota},
where the emergence of these higher excited states could be interpreted
as a manifestation of an effective modulational instability.
Within the effectively finite region determined by the confining
potential, as $g$ is increased, higher ``modulation wavenumbers''
become unstable, leading to the ``quantized'' (associated with the
quantization of wavenumbers in the effectively finite box) cascade
of supercritical pitchfork bifurcations and associated further
destabilizations of the symmetric state.
The degree of phase separation in
Fig.~\ref{exbif} between these bands is computed as
$\Delta h=u_1(x_c)-u_2(x_c)$,
where $x_c$ is the location of the closest density maximum,
irrespective of component, to the trap center.

%
The emergence of high order states can be inferred by observing the
real part of the eigenvalues as the interspecies coupling  is
varied. The eigenvalues collide with the imaginary axis as new
states emerge. Eigenvalue analysis shows that all excited states are
unstable with the exception of the single-hump phase-separated state
(which results from the first pitchfork bifurcation, namely the 1-1
hump state).
These excited states are, however, less unstable than the mixed
state from which they arise. The latter, as shown in Fig.~\ref{exphase},
becomes progressively more unstable, as expected, as further
multi-hump branches arise. Each of the bifurcation points,
for which the same designation as in Fig.~\ref{exbif} is used,
corresponds to a further pair of real eigenvalues appearing
for the mixed branch.
%

%
\begin{figure}[htbp]
\begin{center} \hskip0.15cm
\includegraphics[width=8.5cm]{\rootfig 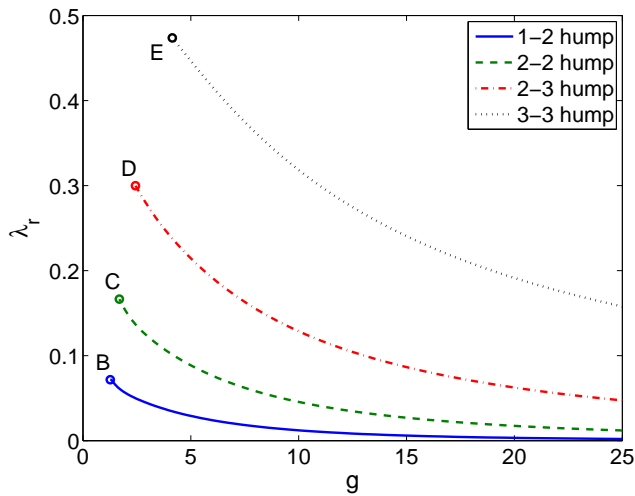}
\caption{(Color online)
Real part of the largest eigenvalues of
the principal phase-separated states
as a function of the interspecies coupling
parameter, $g$, for $\Omega=0.2$.
}
\label{unst_evals}
\end{center}
\end{figure}

As indicated by Fig.~\ref{exbif}, the 1-1 hump
phase separated state is stable even for large values of $g$.
In fact, for $\Omega\ll 1$ and $g\gg 1$, the two
components repel each other so strongly that the center of magnetic trap
becomes a domain wall, where each species abruptly transitions
from near zero atomic density to maximum atomic density.
Apparently, this 1-1 hump phase-separated state is the only
stable state of the system after the mixed state loses
its stability past the bifurcation point A (see Fig.~\ref{exbif}).
Nonetheless, as depicted in Fig.~\ref{unst_evals}, the
additional
 multi-humped excited states are successively created as
$g$ is increased (bifurcation points B, C, D, and E in
Fig.~\ref{exbif}) and are progressively {\em less} unstable as
$g$ increases. Therefore, for sufficiently large $g$,
the instability of higher excited multi-humped states
might be weak enough for these states to be observable
within experimentally accessible times.
It is interesting to note that since these excited
multi-humped states emanate from the (already unstable)
mixed state, they feature a relatively strong
instability close to their bifurcation point.

\begin{figure}[htbp]
\begin{center}
\includegraphics[width=8.5cm]{\rootfig 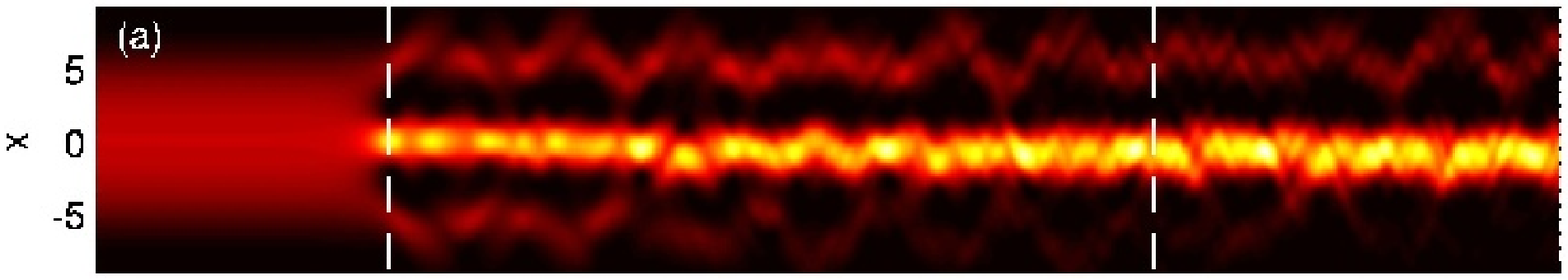}
\vskip-0.1cm
\includegraphics[width=8.5cm]{\rootfig 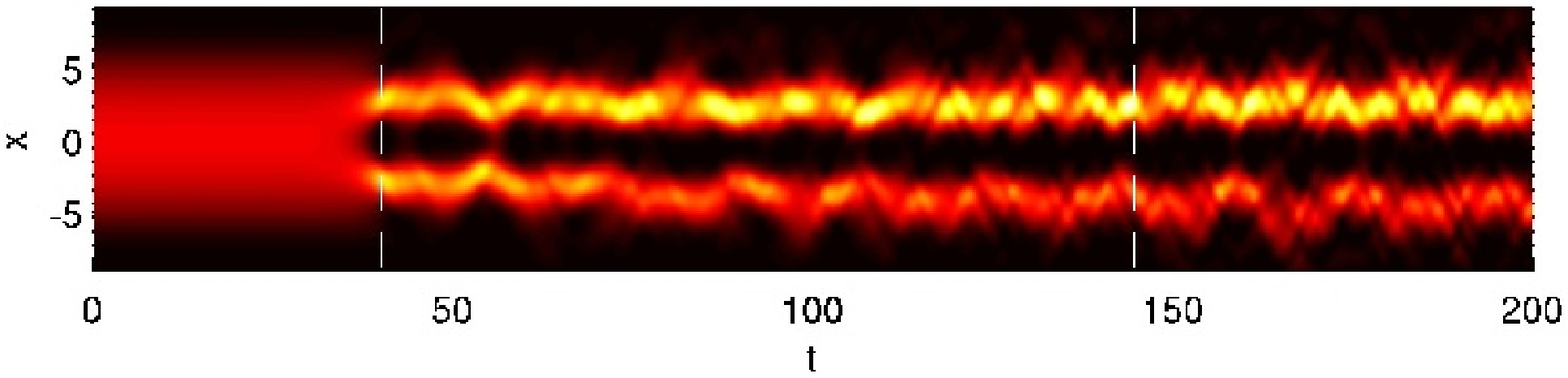}
\hskip-0.6cm
\includegraphics[width=8.2cm]{\rootfig 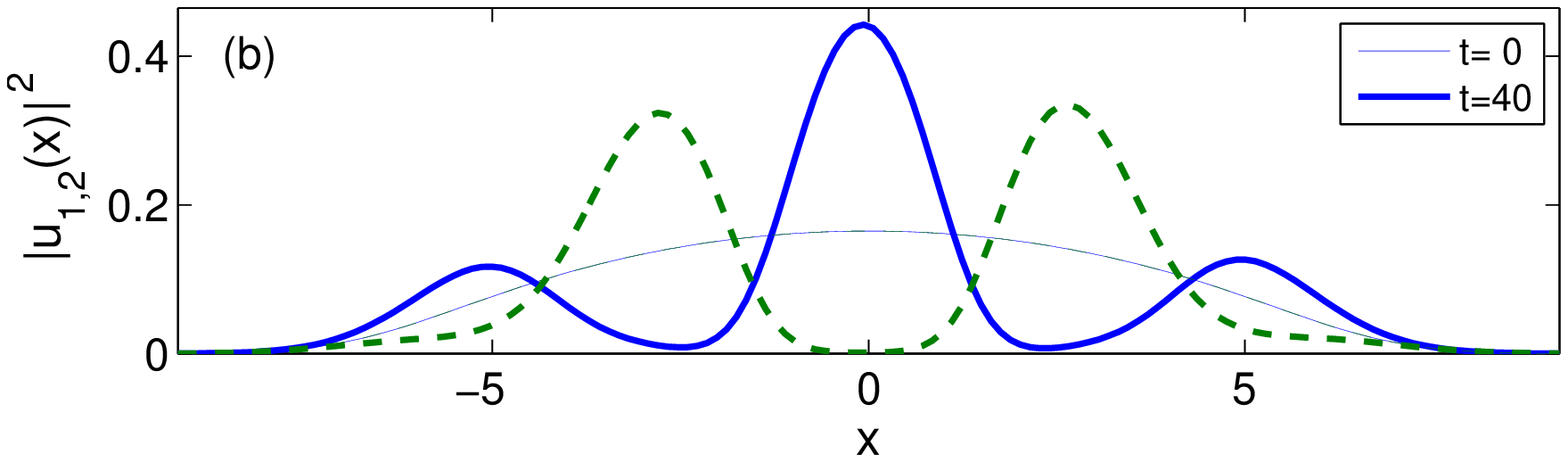}
\includegraphics[width=8.2cm]{\rootfig 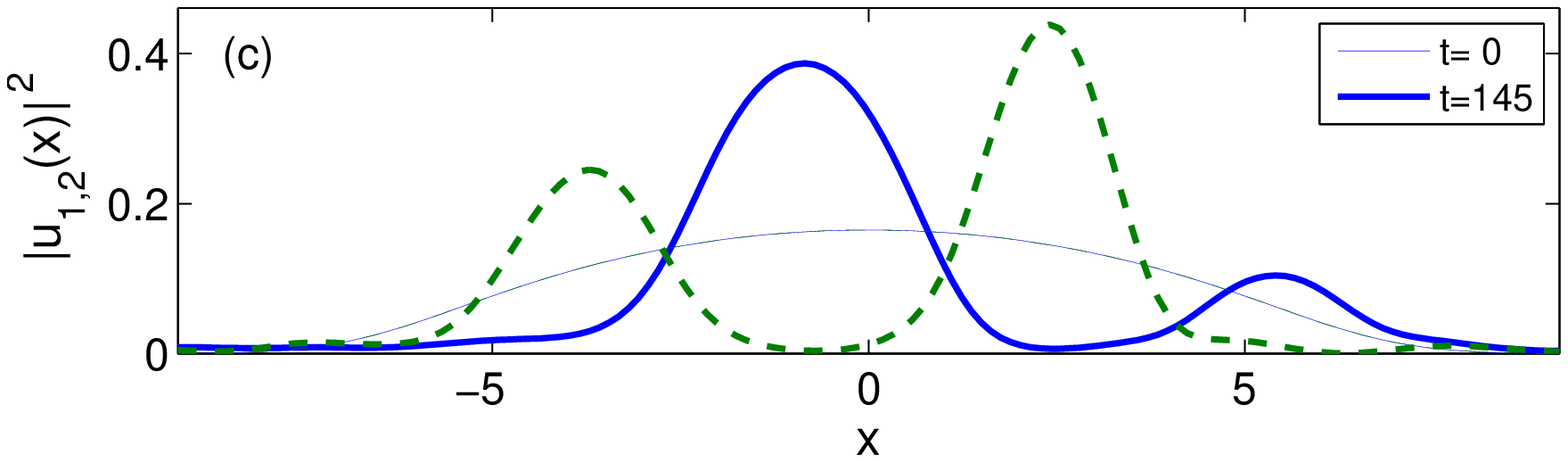}
\includegraphics[width=8.5cm]{\rootfig 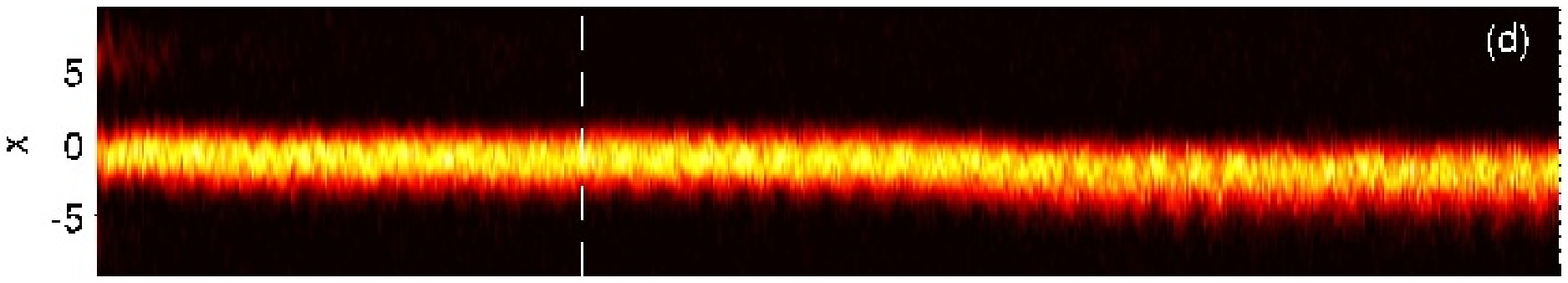}
\vskip-0.1cm
\includegraphics[width=8.5cm]{\rootfig 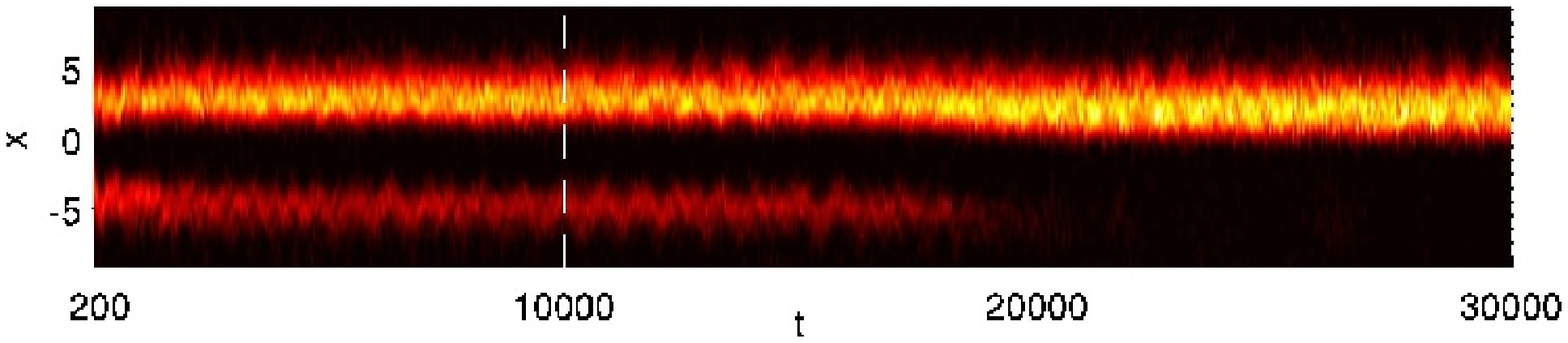}
\hskip-0.6cm
\includegraphics[width=8.2cm]{\rootfig 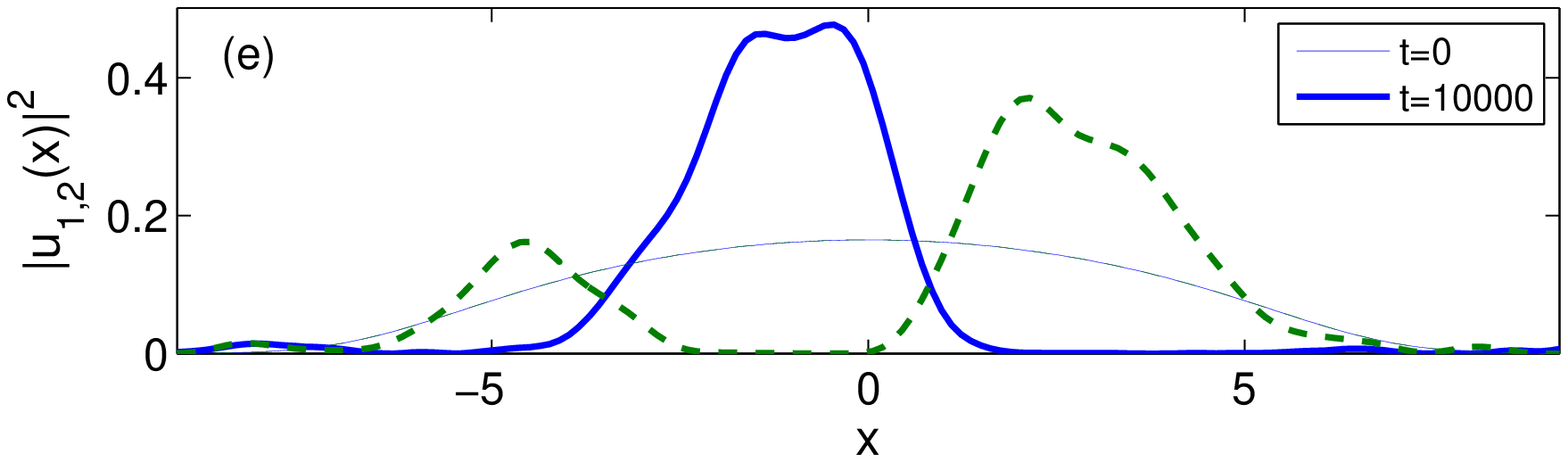}
\caption{(Color online)
Dynamics of the mixed state for $g=5$ and $\Omega=0.2$.
(a) Top and bottom subpanels correspond to the evolution
of the densities for the first and second components, respectively,
after applying an initial spatially random perturbation of
size $1\times 10^{-8}$.
Panels (b) and (c) depict snapshots of the initial
density and at the times indicated in panel (a) by the
white vertical dashed lines.
(d) Long term dynamics showing that the mixed state eventually
approaches a separated state.
(e) Snapshots of the densities at $t=0$ and $t=10,000$
(see white vertical dashed line in panel (d)).
}
\label{u00_dyn}
\end{center}
\end{figure}

\begin{figure}[htbp]
\begin{center}
\includegraphics[width=8.5cm]{\rootfig 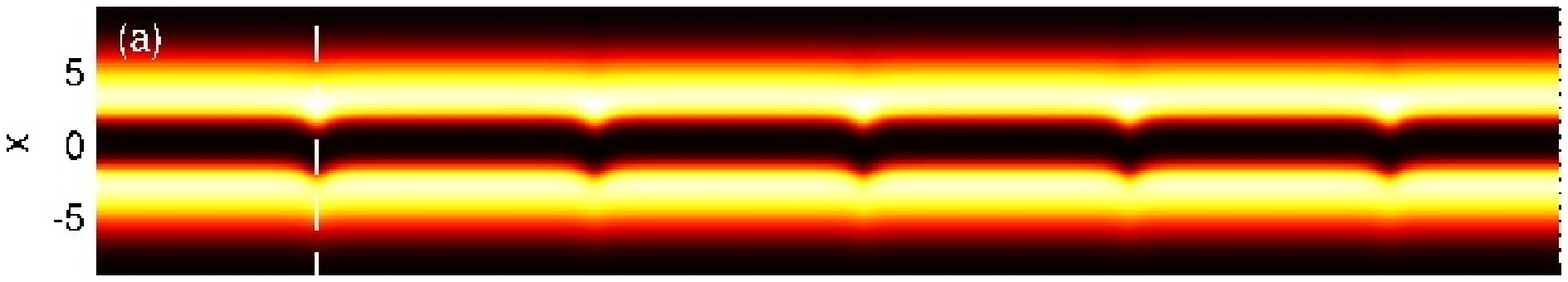}
\vskip-0.1cm
\includegraphics[width=8.5cm]{\rootfig 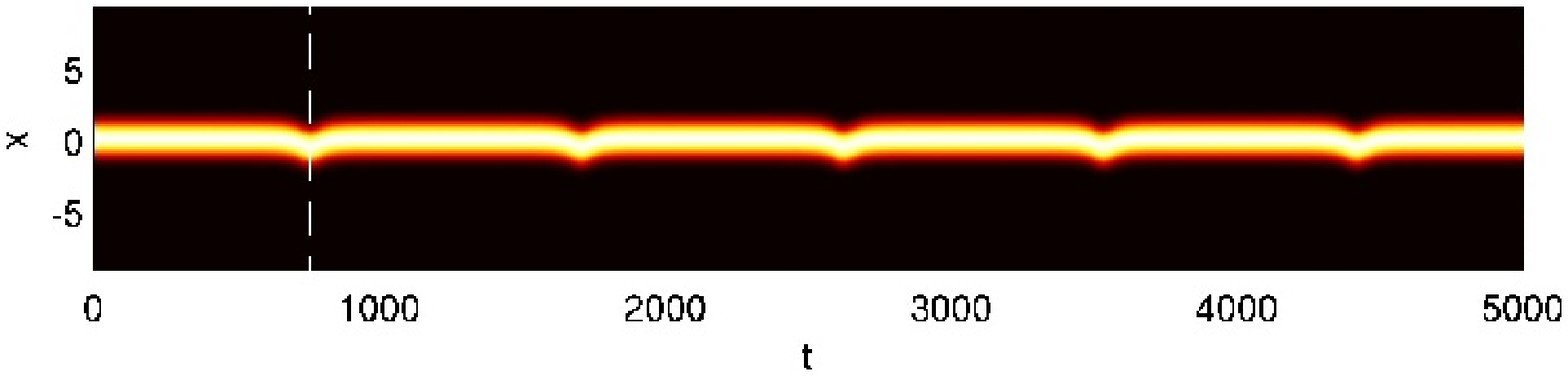}
\hskip-0.6cm
\includegraphics[width=8.2cm]{\rootfig 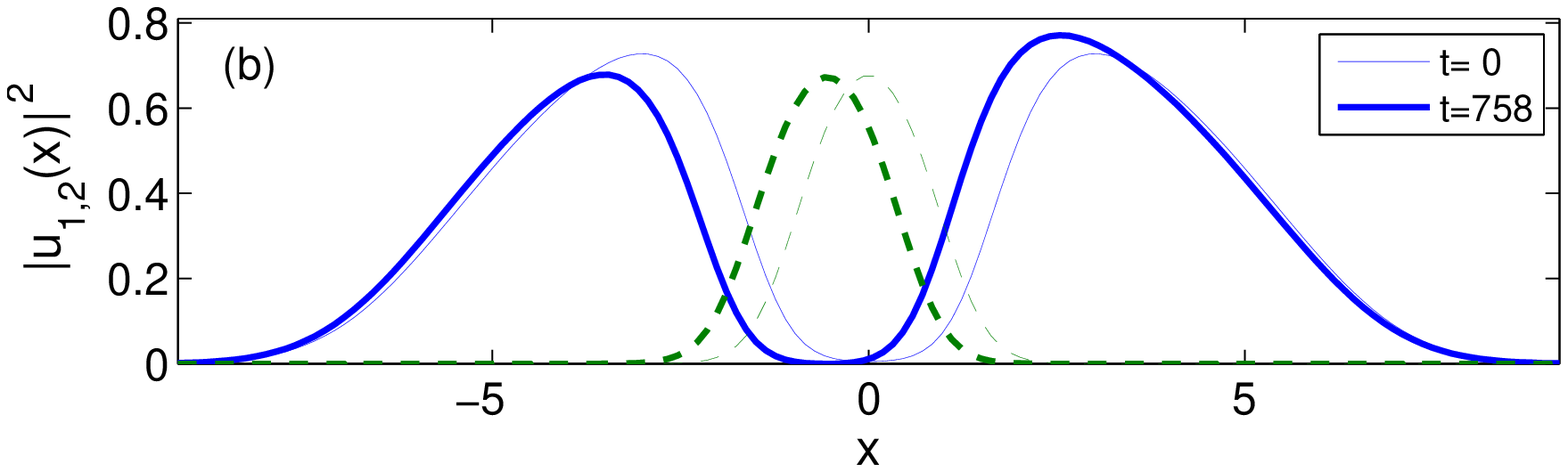}
\includegraphics[width=8.2cm,height=2.6cm]{\rootfig 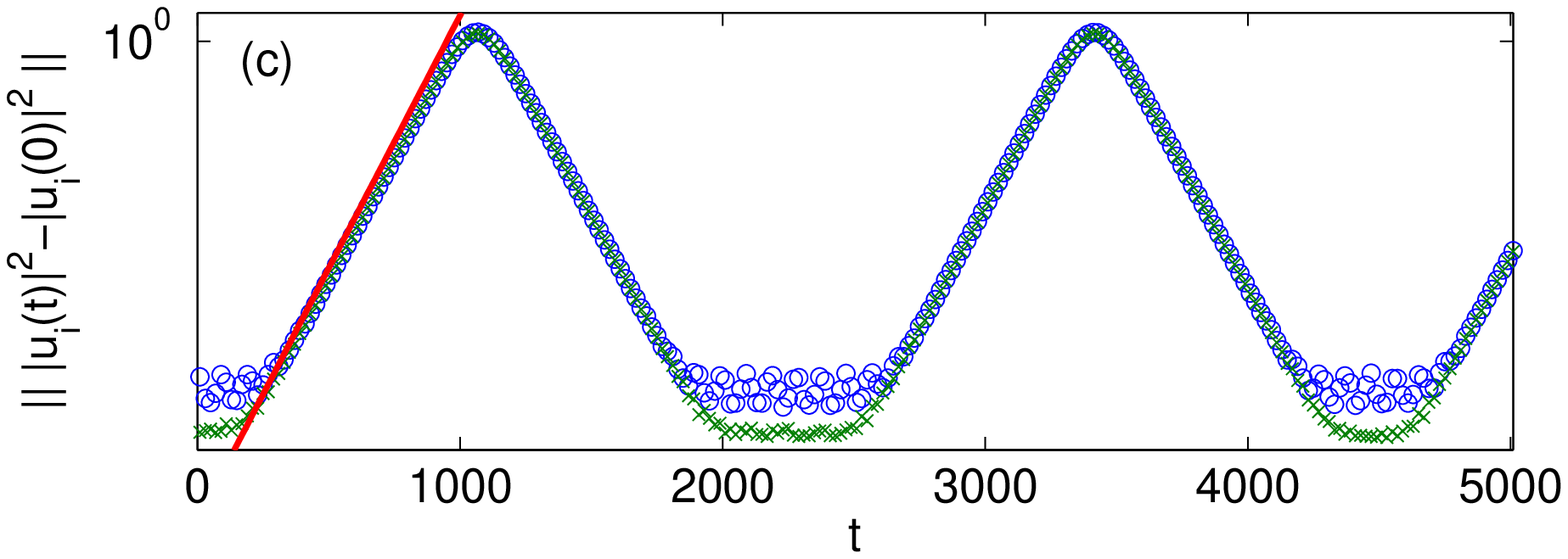}
\caption{(Color online)
Dynamics of the unstable 1-2 hump state for $g=5$ and $\Omega=0.2$.
(a) Same as in Fig.~\ref{u00_dyn}.a.
(b) Snapshots of the densities at $t=0$ and $t=758$ (see white vertical
dashed line in panel (a)).
(c) Growth of the norm of the perturbation
vs.~time (in semi-log plot).
Circles (crosses) correspond to the perturbation of the
first (second) component and the red solid line is the instability
growth obtained from the eigenvalue computation
($\max(\Re(\lambda)))=0.122$).
%
}
\label{u12_dyn}
\end{center}
\end{figure}

\begin{figure}[htbp]
\begin{center}
\includegraphics[width=8.5cm]{\rootfig 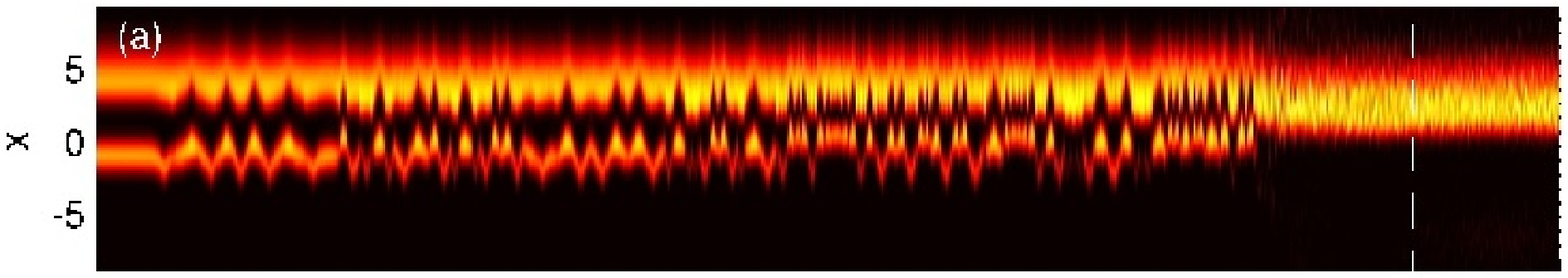}
\vskip-0.1cm
\includegraphics[width=8.5cm]{\rootfig 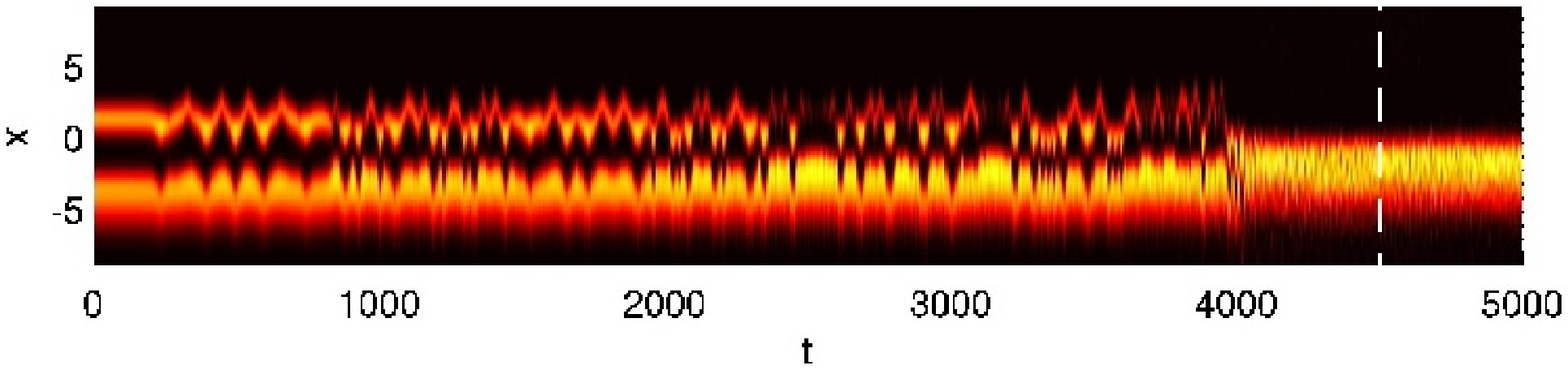}
\hskip-0.6cm
\includegraphics[width=8.2cm]{\rootfig 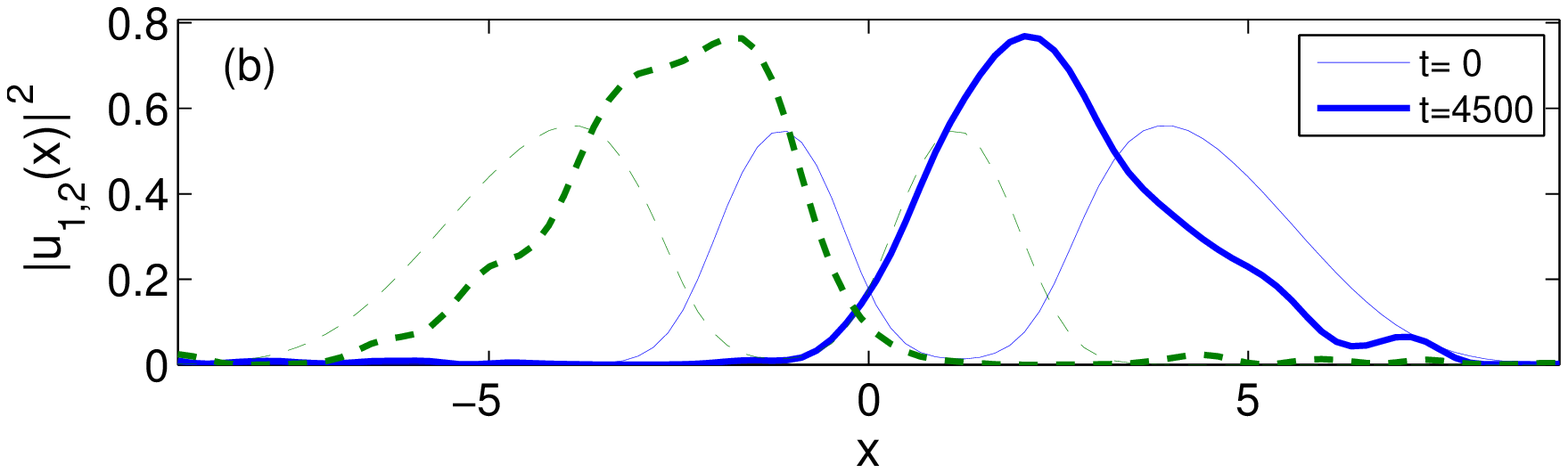}
\caption{(Color online)
Dynamics of the unstable 2-2 hump state for $g=5$ and
$\Omega=0.2$ as in Fig.~\ref{u00_dyn}. The eventual result
of the instability is the formation of a robust and dynamically
stable 1-1 hump state.
}
\label{u22_dyn}
\end{center}
\end{figure}

\begin{figure}[htbp]
\begin{center}
\includegraphics[width=8.5cm]{\rootfig 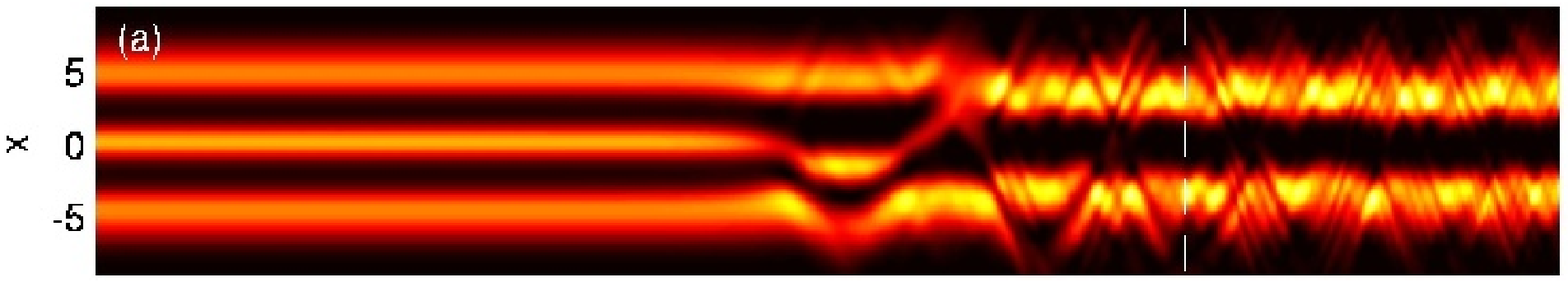}
\vskip-0.1cm
\includegraphics[width=8.5cm]{\rootfig 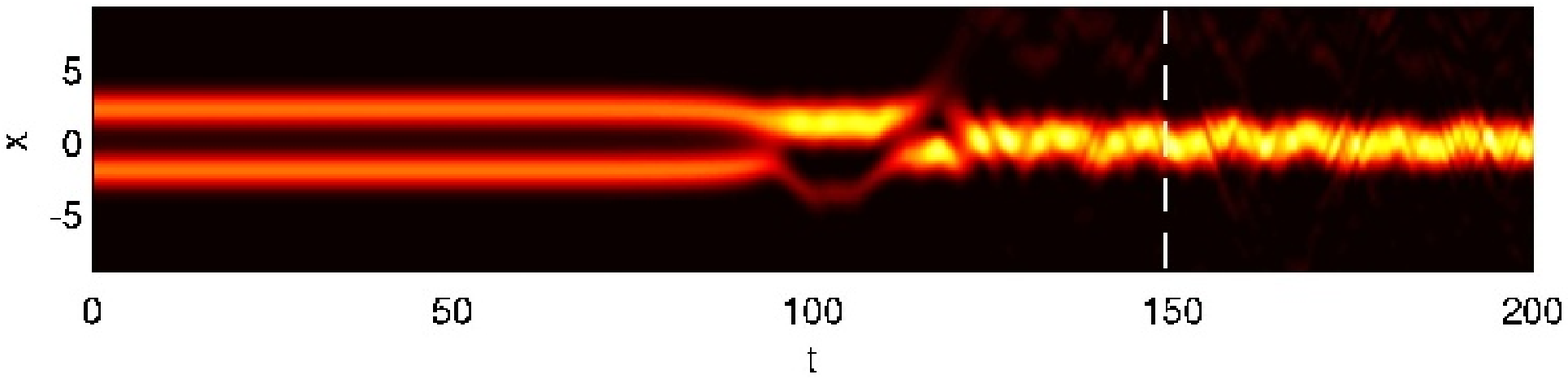}
\hskip-0.6cm
\includegraphics[width=8.2cm]{\rootfig 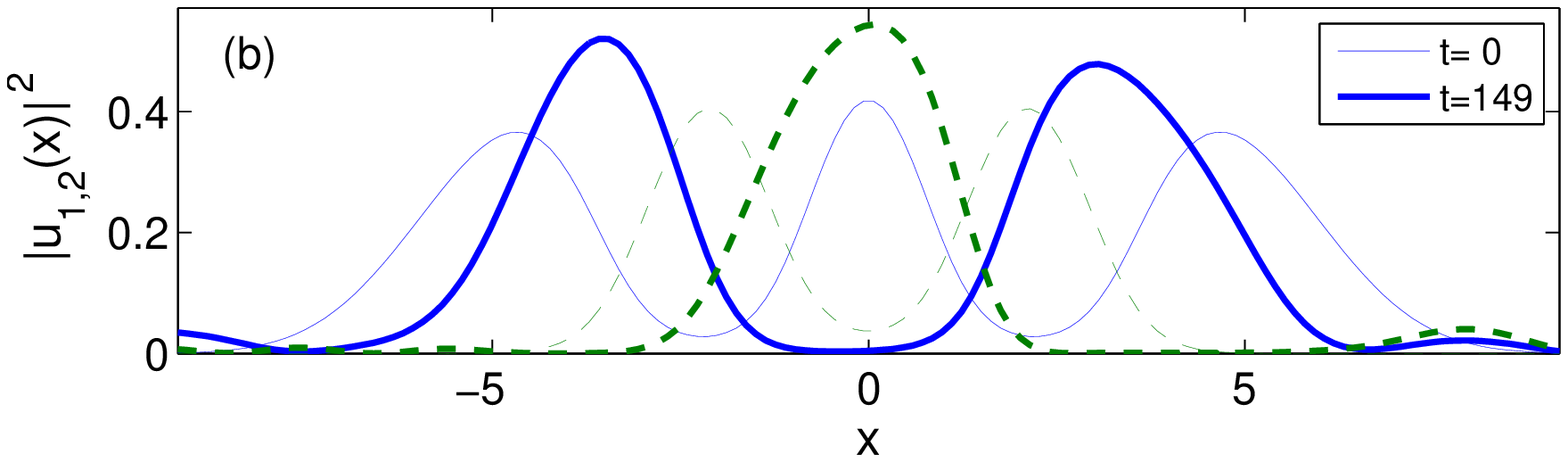}
\includegraphics[width=8.5cm]{\rootfig 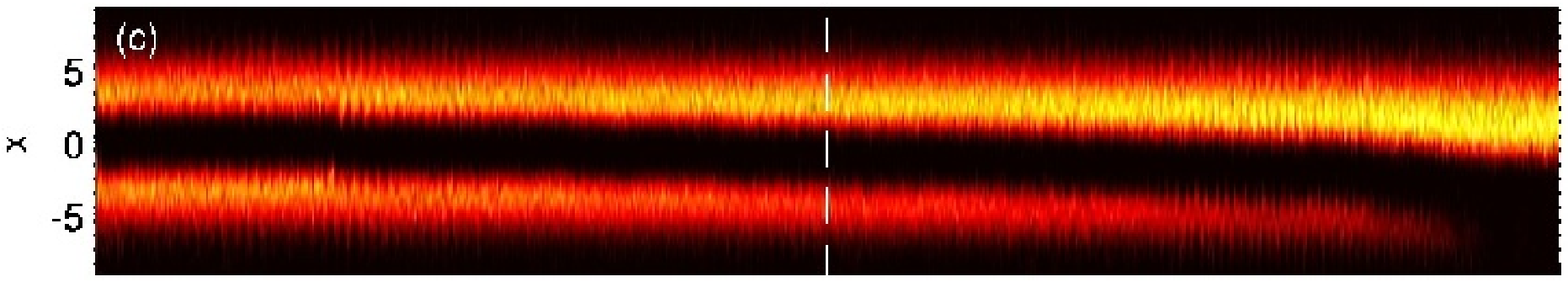}
\vskip-0.1cm
\includegraphics[width=8.5cm]{\rootfig 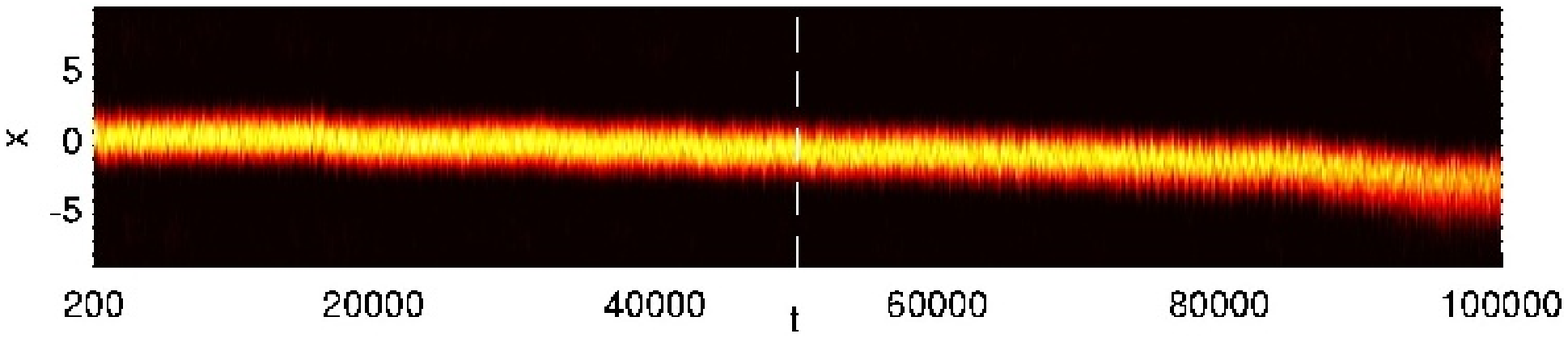}
\hskip-0.6cm
\includegraphics[width=8.2cm]{\rootfig 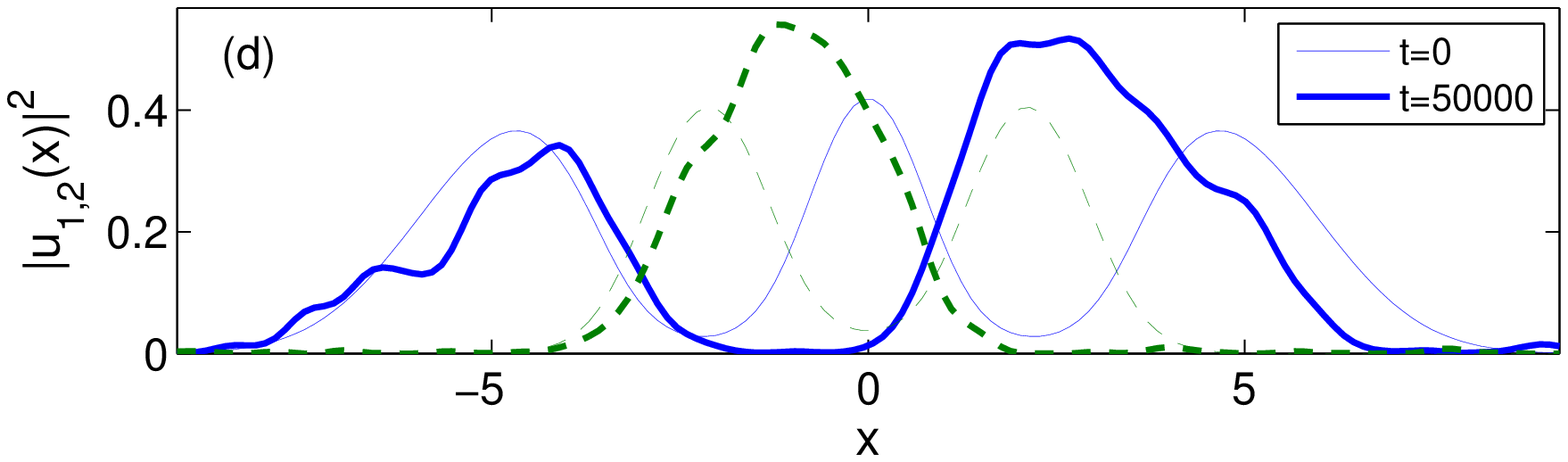}
\caption{(Color online)
Dynamics of the unstable 2-3 hump state for $g=5$ and
$\Omega=0.2$ as in Fig.~\ref{u00_dyn}.
The initial instability leads to a transient
1-2 state that eventually approaches the separated state
for extremely long propagation times ($t>100,000$).
}
\label{u23_dyn}
\end{center}
\end{figure}

\begin{figure}[htbp]
\begin{center}
\includegraphics[width=8.5cm]{\rootfig 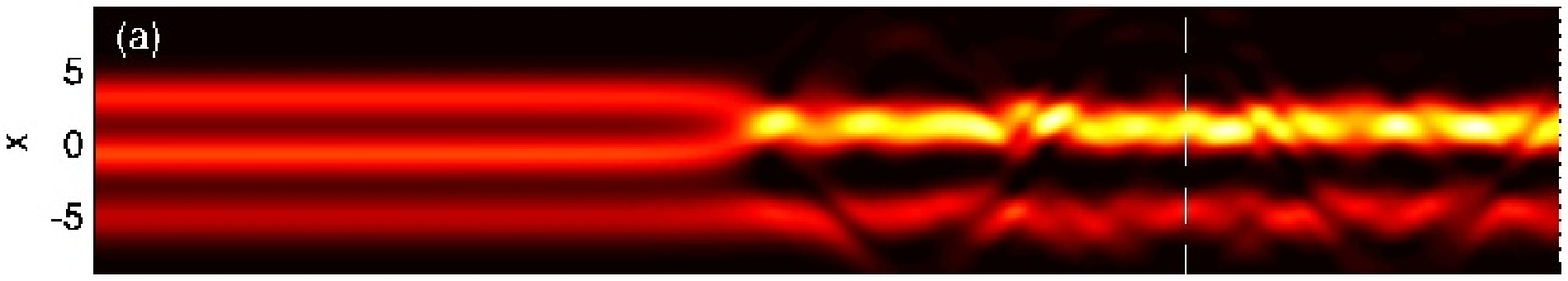}
\vskip-0.1cm
\includegraphics[width=8.5cm]{\rootfig 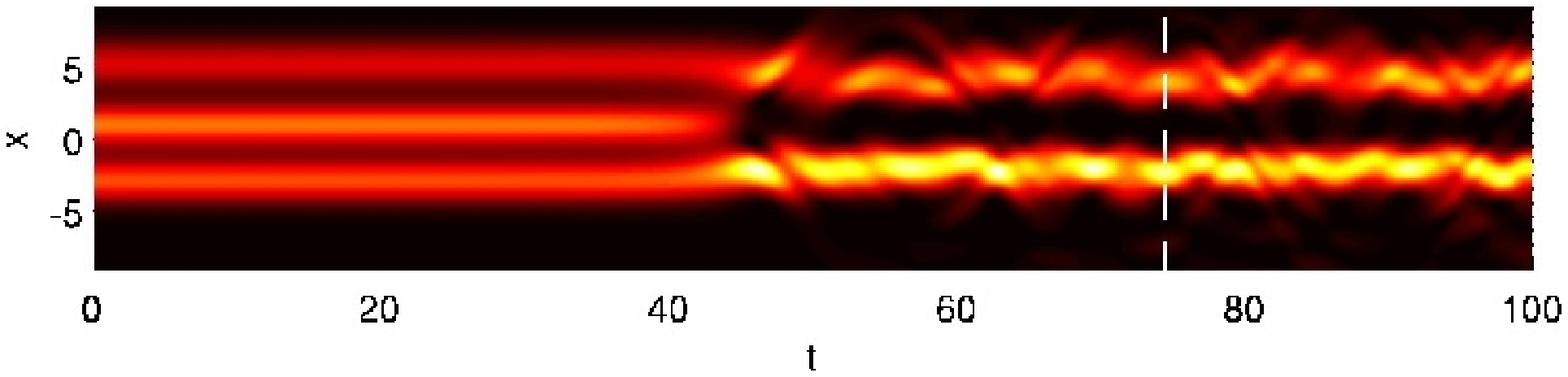}
\hskip-0.6cm
\includegraphics[width=8.2cm]{\rootfig 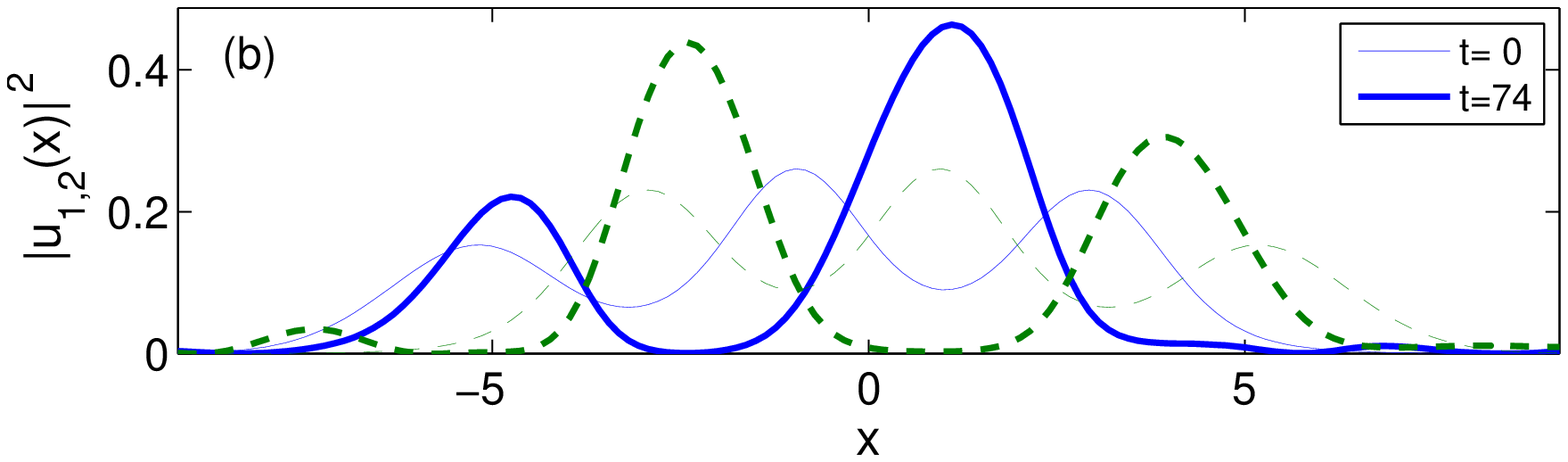}
\includegraphics[width=8.5cm]{\rootfig 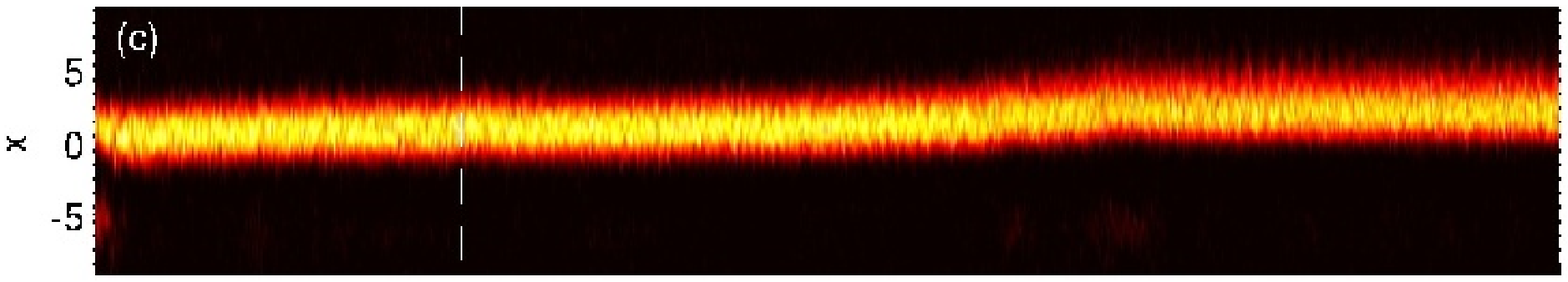}
\vskip-0.1cm
\includegraphics[width=8.5cm]{\rootfig 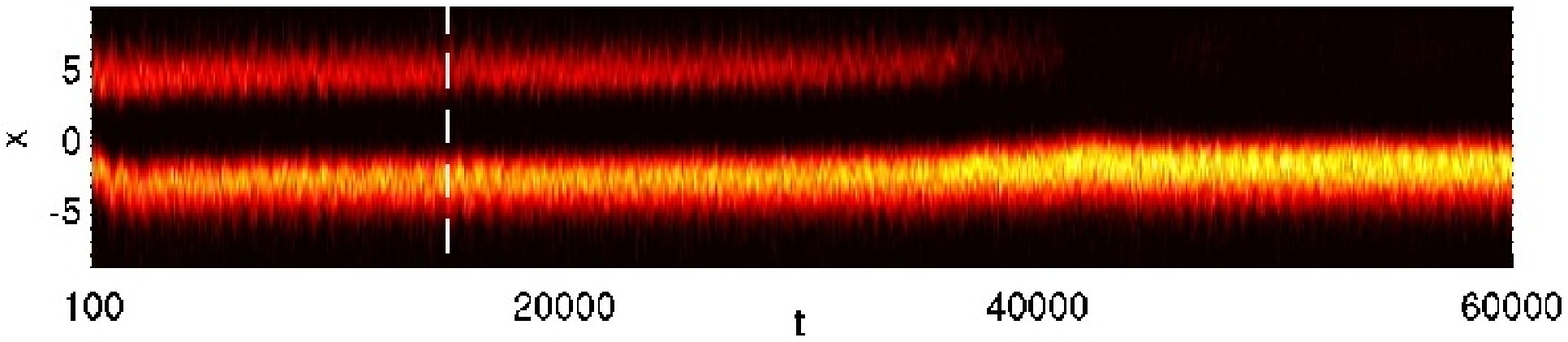}
\hskip-0.6cm
\includegraphics[width=8.2cm]{\rootfig 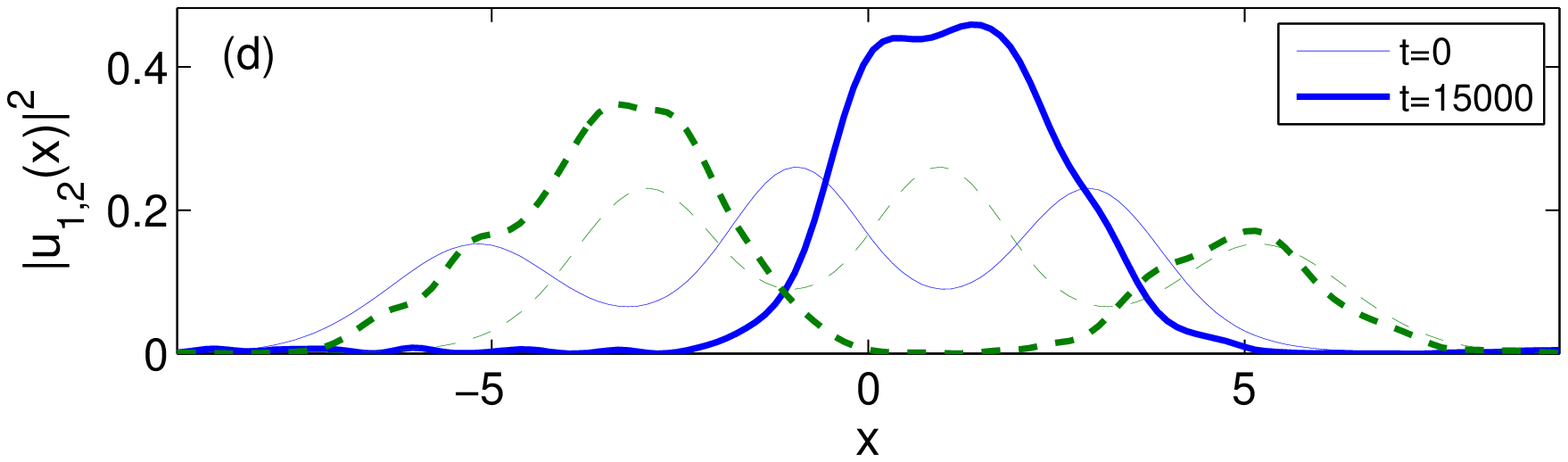}
\caption{(Color online)
Dynamics of the unstable 3-3 hump state for $g=5$ and
$\Omega=0.2$ as in Fig.~\ref{u23_dyn}. The initial
instability dynamics reshapes the waveform into a 2-2 state,
then to long-lived transient a 1-2 state that eventually
approaches the 1-1 state for $t>40,000$.
}
\label{u33_dyn}
\end{center}
\end{figure}

\section{Dynamics of unstable states\label{sec:dynamics}}
In this section we present the dynamics of the
different unstable states that were identified above.
We start by analyzing the destabilization of the
mixed state for a value of $g$ to the right of
the bifurcation point A (see Fig.~\ref{exbif}).
In fact, for all the dynamical destabilization
results presented in this section we chose a
value of $g=5$ and $\Omega=0.2$ that is to the
right of the bifurcation point E (see Fig.~\ref{exbif}).
Therefore, in this regime we have co-existence of
several {\em unstable} multi-hump solutions and the {\em stable}
phase-separated 1-1 hump state. Other parameter regions
(not shown here) gave qualitatively the same results.

In Fig.~\ref{u00_dyn} we show the destabilization of
the mixed state. As it can be observed from the figure,
the mixed state suffers an initial modulational instability
that becomes apparent for $t>35$ that seeds a
highly perturbed 2-3 hump solution [see panel (b)].
Since this new
solution is also unstable for the chosen parameter values,
it is rapidly converted into a relatively long
lived 1-2 hump state [see panel (c)].
Nonetheless, as it is clear from the long term dynamics
presented in panel (d), the 1-2 hump state, being
unstable, eventually ``decays'' to the separated
(1-1 hump) state. We use here the term ``decaying''
in quotes since our system is conservative (no dissipation)
and thus there is no real decay.

In Fig.~\ref{u12_dyn} we show the destabilization
of the unstable 1-2 hump state. As it is obvious from the
figure, although the instability eigenvalue
is sizeable [$\lambda_r=0.122$, see panel (c)]), the initially
weak perturbation does not lead to the destruction
of the 1-2 hump state. Instead, this unstable state
just momentarily ``jumps'' to the left
(or to the right, results not shown here), see panel (b),
for a short period of time (becoming slightly asymmetric)
and then comes back close
to the 1-2 hump (symmetric) steady state configuration.
This indicates that, for this parameter combination,
the 1-2 hump state is a saddle fixed point.
Thus, the orbit remains close to the steady 1-2
hump state and it is eventually ``kicked out''
along the unstable manifold. Then, it performs a
quick excursion and ``comes back''
through the stable manifold.

In Fig.~\ref{u22_dyn} we depict the destabilization dynamics
of the 2-2 hump state. As it is evident from the figure, when
compared to Fig.~\ref{u12_dyn}, the destabilization
happens earlier ($t \approx 200$)
since the 2-2 hump state is more unstable
than the 1-2 hump state (see Fig.~\ref{unst_evals}).
As the dynamics of the perturbed 2-2 hump state evolves,
it progressively favors phase separation until
eventually the system rearranges itself into a
1-1 hump state at about $t\approx 4000$. Since
the 1-1 hump is stable, this resulting configuration
is sustained thereafter.

A similar scenario (fast initial destabilization, slow transient
stage and eventual settling into the stable phase-separated state)
is observed when following the
dynamical destabilization of the 2-3 hump state.
As it can be observed from Fig.~\ref{u23_dyn},
the relevant configuration destabilizes around $t\approx 90$
(earlier than its 1-2 and 2-2 hump counterparts given
its larger instability growth rate [see Fig.\ref{unst_evals}]).
The dynamics goes through a transient 1-2 state and eventually
settles into a highly perturbed separated state that is preserved
thereafter as was the case also for the mixed state dynamics
(see Fig.~\ref{u00_dyn}) and the 2-2 state
(see Fig.~\ref{u22_dyn}) presented above.

Finally, in Fig.~\ref{u33_dyn} we present the dynamical
destabilization for the 3-3 hump state. Again, this
state destabilizes even faster ($t=40$) than the previous
waveforms because of its higher instability eigenvalue
(see Fig.~\ref{unst_evals}).
In this case, the 3-3 hump state first destabilizes into a
highly perturbed 2-2 state. Since the 2-2 hump state is
unstable, the ensuing dynamics
results into a separated 1-2 hump state after $t>100$
which, in turn, eventually settles to a highly perturbed separated
state.

\begin{figure}[htbp]
\begin{center}
\includegraphics[width=8.5cm]{\rootfig 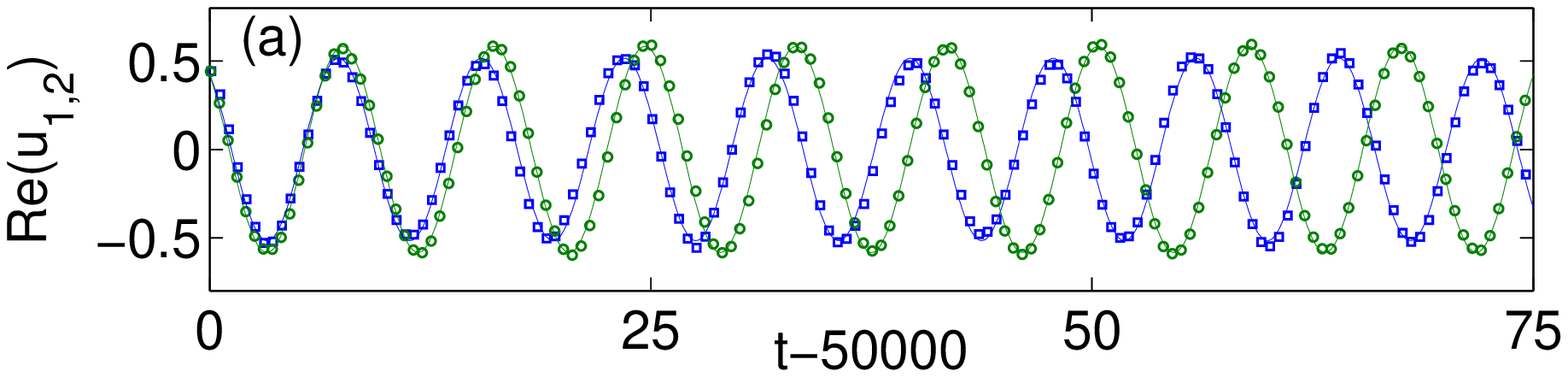}
\includegraphics[width=8.5cm]{\rootfig 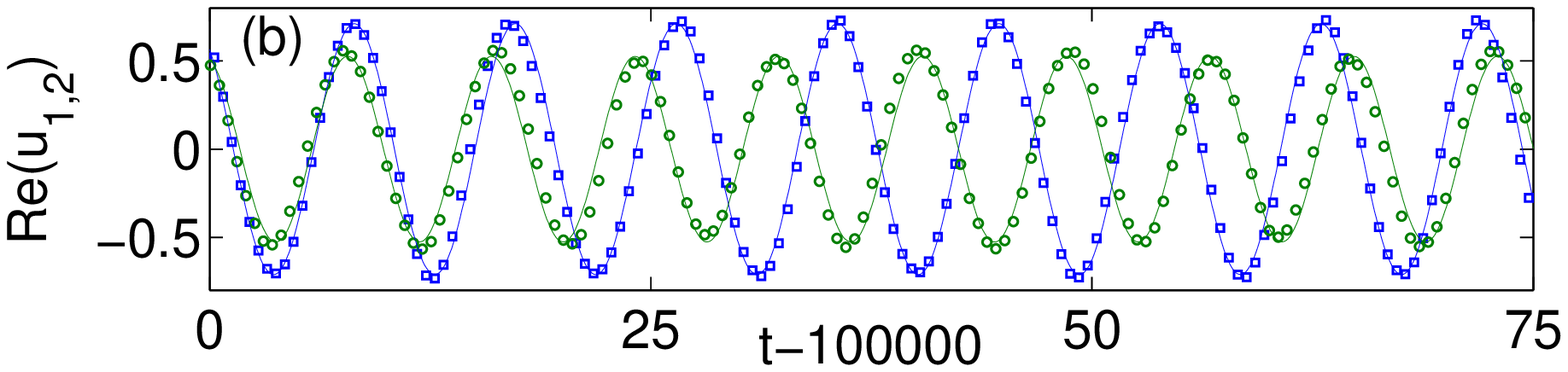}
\caption{(Color online)
Dynamics of the real part of the wave functions for the 2-3 hump
state of Fig.~\ref{u23_dyn} for $g=5$ and $\Omega=0.2$.
Panels (a) and (b) correspond, respectively, to times centered about
$t=50,000$ and $t=100,000$. The blue square (green circles) correspond
to the first (second) component. The continuous lines correspond
to the best sinusoidal fit from which the local chemical potential
can be extracted.
}
\label{u23_diff_mu}
\end{center}
\end{figure}

\begin{figure}[htbp]
\begin{center}
\includegraphics[width=8.5cm]{\rootfig 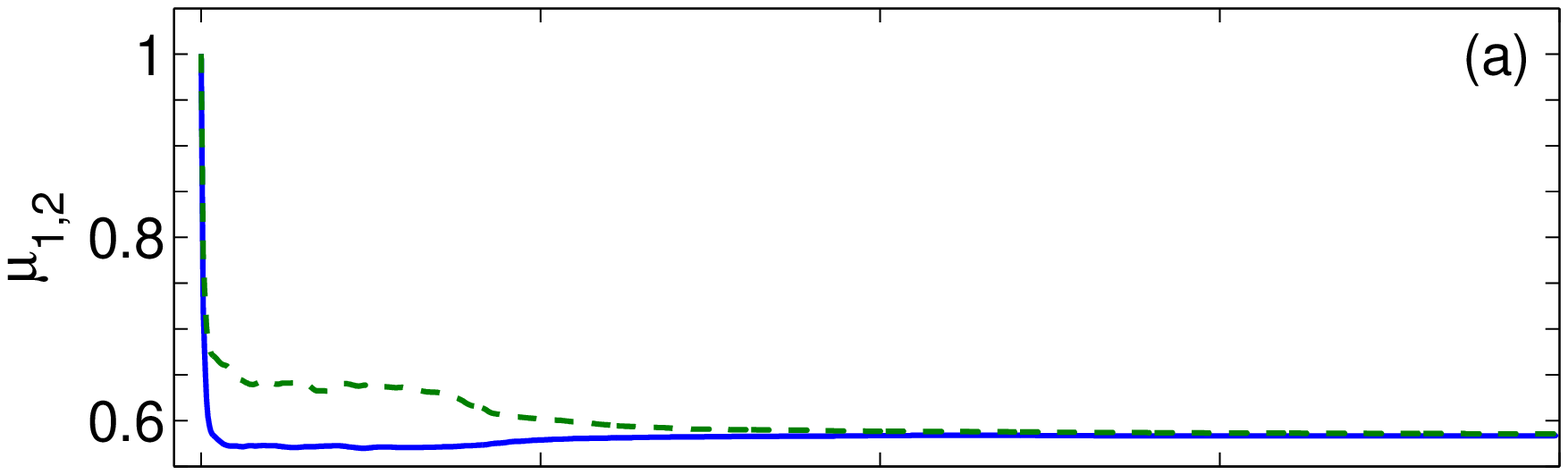}
\\
\includegraphics[width=8.5cm]{\rootfig 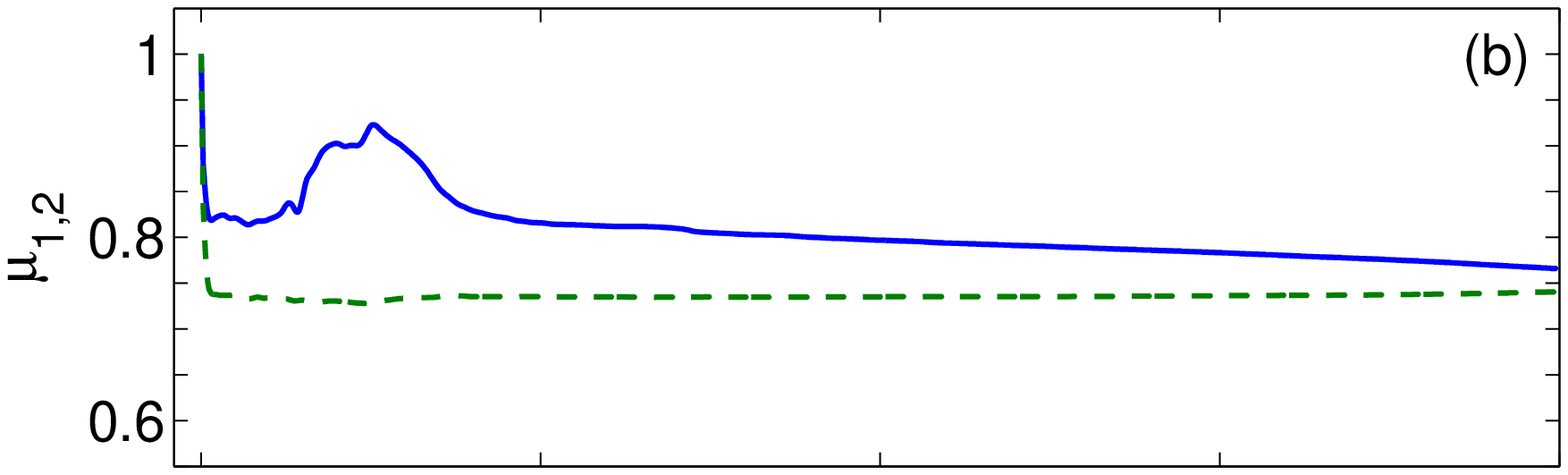}
\\
\includegraphics[width=8.5cm]{\rootfig 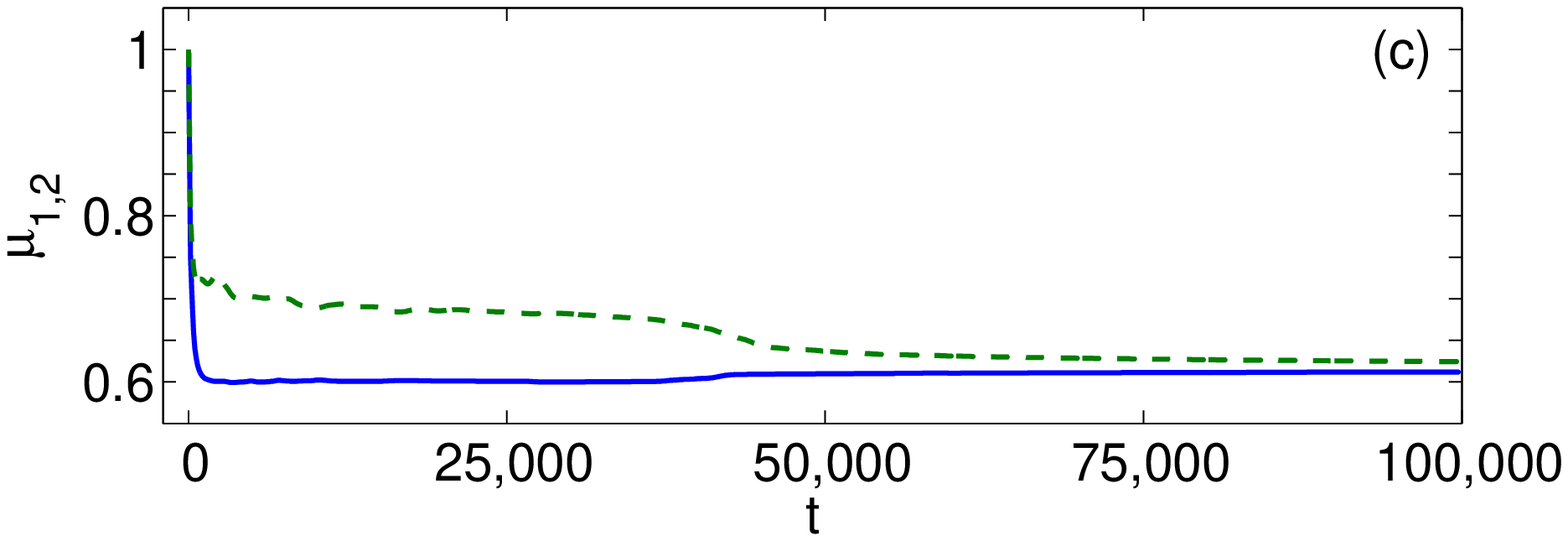}
\caption{(Color online)
Local chemical potential at the location of maximum
density for each component. Blue solid line corresponds to
the first component while the green dashed line corresponds to
the second component. The different panels
correspond to the evolution of the local
chemical potentials for the steady states:
(a) mixed state (cf.~Fig.~\ref{u00_dyn}),
(b) 2-3 hump state (cf.~Fig.~\ref{u23_dyn}),
and
(c) 3-3 hump states (cf.~Fig.~\ref{u33_dyn}).
}
\label{u00_23_33_long_mu}
\end{center}
\end{figure}

\begin{figure*}[htbp]
\begin{center}
\includegraphics[width=13cm,height=10cm]{\rootfig 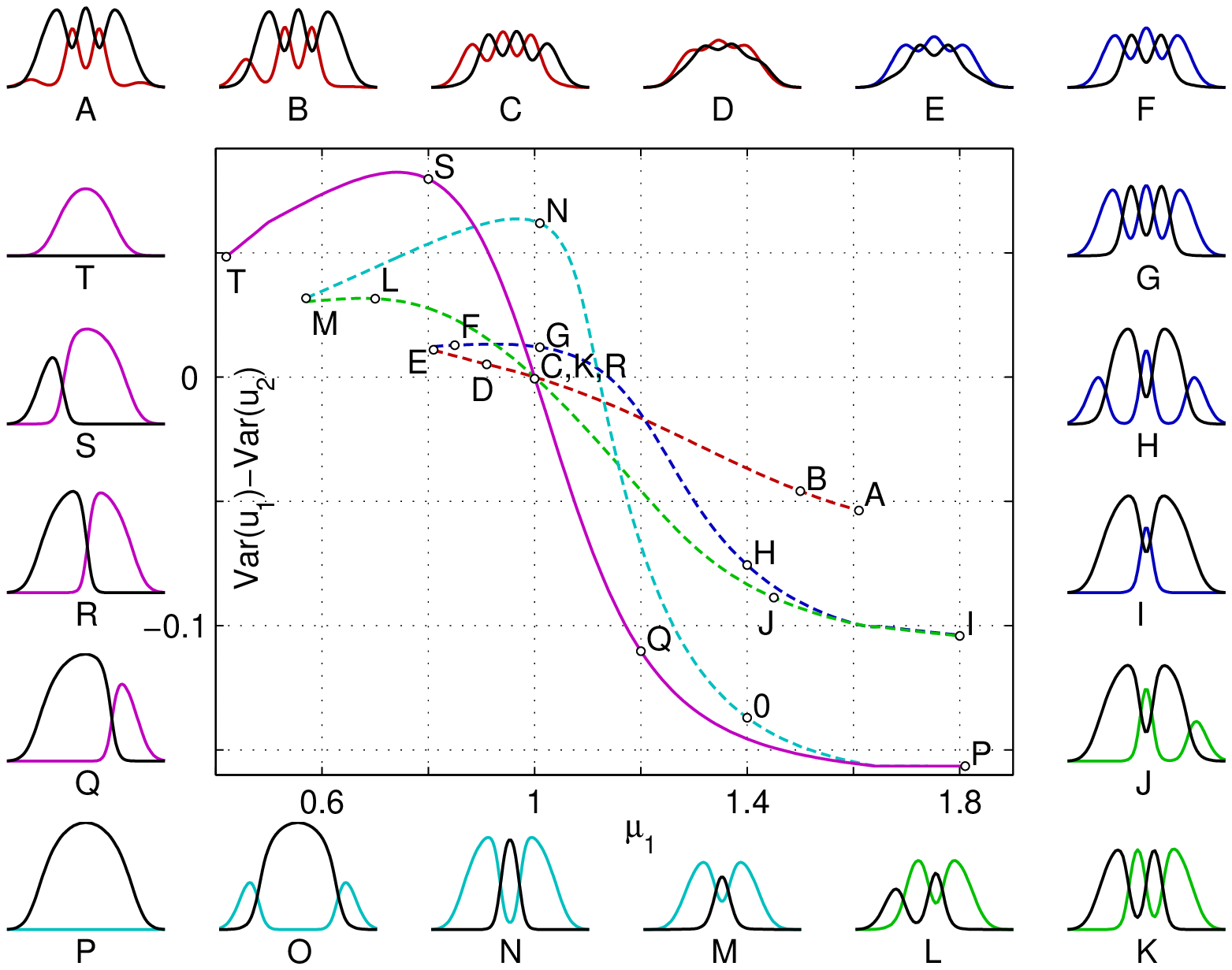}
\vspace{-0.2cm}
\caption{(Color online)
Bifurcation diagram of asymmetric states for constant
$\mu_2=1$ as a function of $\mu_1$. The vertical axis
corresponds to the difference of the variance of the
steady state configurations for both components
for $g=5$ and $\Omega=0.2$. Stable (unstable) solutions
branches are depicted with black solid (dashed) lines.
Typical solutions for each branch are depicted in the
surrounding insets.}
\label{mu_bif1}
\end{center}
\end{figure*}


\section{Existence and stability of asymmetric states\label{sec:asym}}
It is interesting to observe in the above
computations that the transient 1-2 hump
state emanating from the destabilization of the
mixed state (see Figs.~\ref{u00_dyn}) or from  higher
order states (see Figs.~\ref{u23_dyn} and \ref{u33_dyn})
has a relatively long life span. However, being this
state also unstable, it eventually
tends to a perturbed separated (i.e., 1-1 hump) state.
The process of converting a 1-2 state into the
separated state involves the effective shift of
mass from one of the two humps (in the component with
two humps) to the other one until all the mass
is ``swallowed up'' by one hump resulting in a
1-1 hump. For example, as it can be observed in panel (d)
of Fig.~\ref{u23_dyn}, at $t=50,000$ the right hump of
the first component has more mass than its left
hump. Since the chemical potential (i.e., rotation
frequency of the wave function in the complex plane)
is closely related to the mass of the condensate,
it is possible to follow the local change in mass
by following the {\em local} chemical potential.
In Fig.~\ref{u23_diff_mu} we depict the oscillations
of the real part of the wave functions at the location
of the {\em maximal} density (see squares and
circles in the figure). We then fit a sinusoidal
curve (see solid lines in the figure) through the data
to obtain the oscillation frequency. As it is obvious
from the figure, although both components started with
the 2-3 hump steady state with the {\em same} chemical
potential $\mu_1=\mu_2=\mu=1$,
the two components oscillate at
different rates. In fact, at $t=50,000$ the local
chemical potentials for each species is, respectively,
$\mu_1=0.776$ and $\mu_2=0.734$, and at $t=100,000$
we have $\mu_1=0.688$ and $\mu_2=0.771$.
We systematically extracted the local chemical potentials
at the location of the maximum density for the
mixed state, 2-3 hump state and 3-3 hump state
and depict them in Fig.~\ref{u00_23_33_long_mu}.
As it can be observed from the figure, both
chemical potentials start at $\mu_1=\mu_2=1$ but
rapidly drop when the initial instability of
the steady state develops for the three cases.
Then, the chemical potentials slowly drift until
they acquire values very close to each other at
around the time when the dynamics settles to a
perturbed phase separated state.

\begin{figure}[htbp]
\begin{center}
\includegraphics[width=7cm,height=4cm]{\rootfig 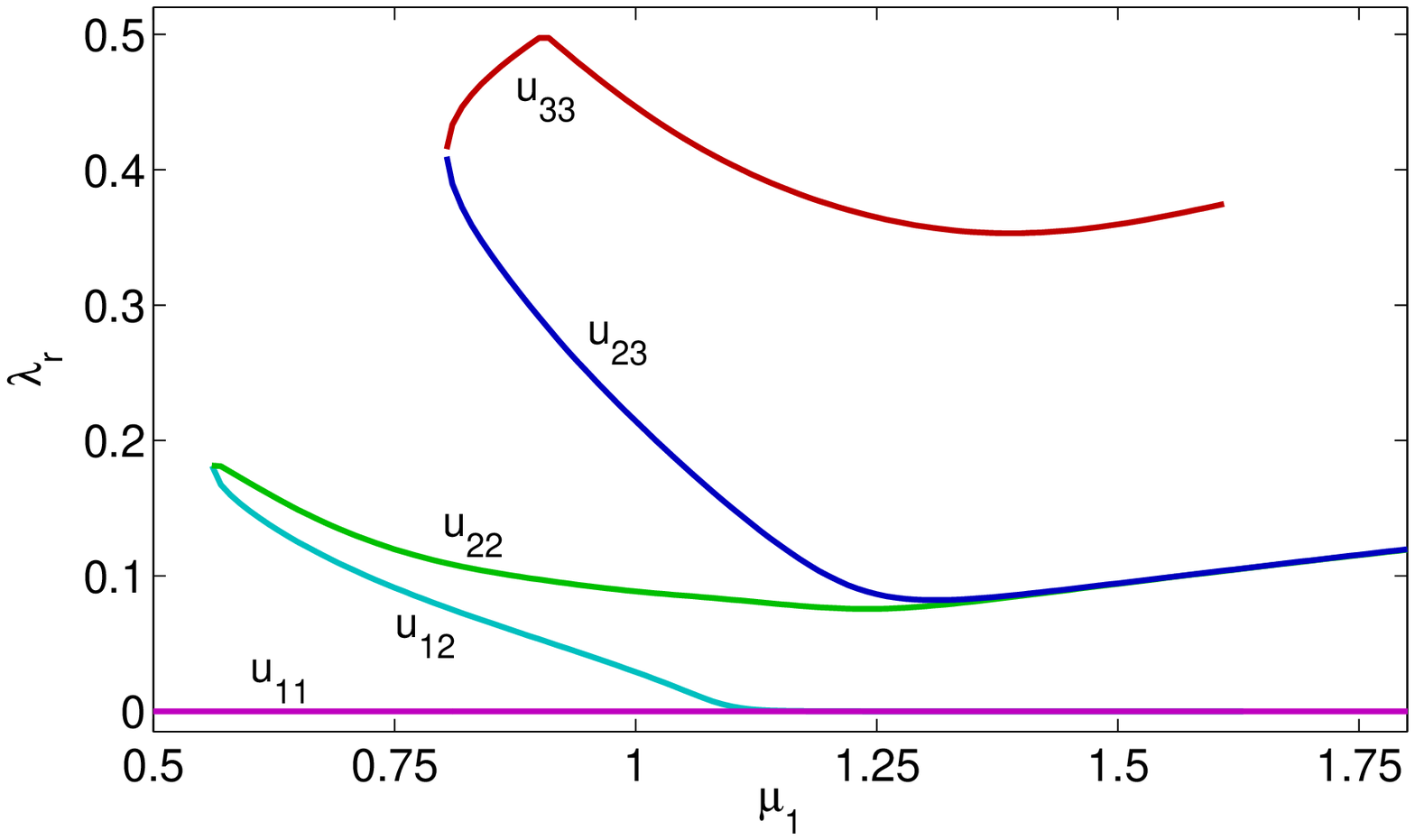}
\\
\includegraphics[width=7cm,height=4cm]{\rootfig 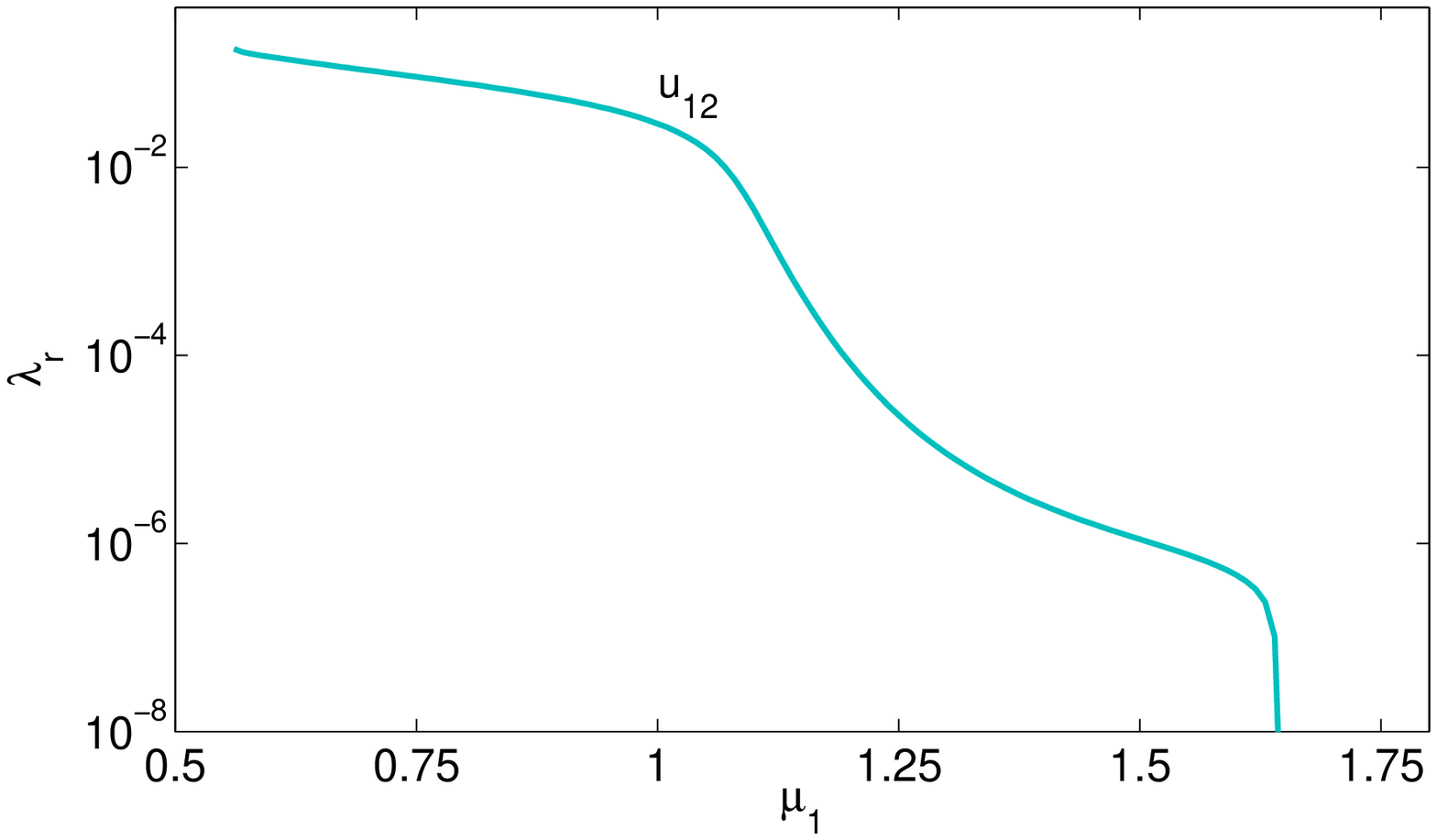}
\vspace{-0.2cm}
\caption{(Color online)
Real part of largest eigenvalue for the asymmetric
mixed states as a function of $\mu_1$ with $\mu_2=1$,
$g=5$ and $\Omega=0.2$.
The bottom panel corresponds to the eigenvalue
for the 1-2 hump state plotted in logarithmic scale.
}
\label{mu1_stab}
\end{center}
\end{figure}


The fact that during the dynamical evolution of the
unstable higher order states the local chemical
potentials vary naturally prompts the question
of existence and stability of steady states
with different chemical potentials between the
species. Up to this point all the steady states we
analyzed supposed that both components had the same
chemical potential. Namely, the evolution for
each components is
\begin{eqnarray}
\nonumber
u_1(x,t) &=& w_1(x) \, e^{-i \mu_1 t}\\[2.0ex]
\nonumber
u_2(x,t) &=& w_2(x) \, e^{-i \mu_2 t},
\end{eqnarray}
where the steady state profiles are $w_{1,2}(x)$ and
$\mu_1=\mu_2=1$. We now relax this symmetric assumption
and look for steady states with different chemical
potentials between the components ($\mu_1\not=\mu_2$).
A full bifurcation diagram as a function of
$\mu_1$, for fixed $\mu_2=1$, is depicted in Fig.~\ref{mu_bif1}.
In this diagram we use in the vertical axis the difference
in the variance of the steady state distribution between
the two components. The different bifurcation branches correspond
to the particular asymmetric states depicted in the
surrounding insets. This diagram provides an account
of how the different solutions bifurcate from each other.
It is very interesting to follow the bifurcation path
of all the states labeled from A to T during which all
the higher order states are browsed continuously.
Due to symmetry,
a similar scenario is present when one keeps $\mu_1=1$
fixed and varies $\mu_2$ (results not shown here).

The stability in the bifurcation diagram depicted
in Fig.~\ref{mu_bif1}
is denoted,
as before, with a solid line for stable states and a
dashed line for unstable states. As it can be noticed
from the figures, as it was the case for symmetric
states, all the asymmetric states are unstable except
the mixed state. This is due to the fact that the
interspecies coupling for these diagrams was chosen
as $g=5$, i.e., high enough so that the mixed state
in unstable.
In fact, in Fig.~\ref{mu1_stab}
we show the largest real part of the eigenvalues
for the different asymmetric states depicted in
the bifurcation diagram in Fig.~\ref{mu_bif1}.
As it was observed for the
symmetric case, the higher the order of the
asymmetric state the more unstable it becomes.
It is crucial to note that the 1-2 hump state
seems to get stabilized for $\mu_2=1$ and
$\mu_1>1.12$ (see top panel of Fig.~\ref{mu1_stab}).
However, after close inspection, see the
bottom panel in Fig.~\ref{mu1_stab},
the real part of the eigenvalue
for the 1-2 hump state {\em never} vanishes but
instead becomes extremely small (between $10^{-8}$
and $10^{-2}$) until the branch disappears
(at $\mu_1 \approx 1.65$ for $\mu_2=1$).
%
This very weak instability explains the fact why
the transient 1-2 hump state appears to be sustained for very
long times in the evolutions depicted in
Figs.~\ref{u00_dyn}, \ref{u23_dyn}, and
\ref{u33_dyn}.

\section{Conclusions\label{sec:conclu}}
In the present work, we have analyzed the emergence of
non-topological,
phase-separated states in the immiscible regime
out of mixed ones in the miscible regimes, as a natural
parameter of the system (namely the interspecies interaction strength)
was varied. Our analysis was presented for the case of
magnetically trapped two-component
Bose-Einstein condensates i.e., two incoherently coupled NLS
equations with a parabolic potential. Using a variational approach,
we obtain a reduced model describing the statics, stability
and dynamics of each condensate cloud. We are able
to elucidate the miscibility boundary for the interspecies
coupling parameter as the strength of the magnetic trap
(and/or the chemical potential of the system) is varied.
The approach is also capable of accurately
capturing the spatial oscillations of the clouds about the
stable stationary states (both for mixed and for phase-separated states).
In particular, for relatively
small interspecies coupling, the two condensed clouds
do not phase-separate (mixed state) giving rise, if the BEC clouds
are initially displaced with respect to each other, to
oscillations through each other. On the
other hand, for relatively large interspecies coupling,
the two clouds form a stable phase-separated state which can
entertain oscillations about the equilibrium separation between
the components.
A further dynamical reduction allows to understand this
behavior more intuitively based on an effective potential that undergoes
a bifurcation from a single well (mixed state) to a double well
(phase-separated state) form as the interspecies
coupling parameter is increased.
We also describe the bifurcation scenario of higher order
phase-separated states, as the
interspecies coupling parameter is increased.
We observe that several (interwoven
between the two components) bands of density modulations
progressively arise out of the mixed states giving rise
to higher excited states.
%
Among all the phase-separated states, only the first
excited one with one hump in each component is found to
be dynamically stable for all values of the interspecies
interaction strength past its bifurcation point.
We furthered our analysis by studying the existence
and stability of asymmetric states for which the
chemical potentials for each species is different.
We found, similar to the symmetric case of equal
chemical potentials, that the only stable steady
state for high enough interspecies coupling is the
separated state. Nonetheless, we found that in
some regimes, the (asymmetric variant of the)
state with one band in one
component surrounded by two bands of the other
component (the 1-2 hump state) can have an
extremely weak instability and thus facilitating
its potential observability in numerical experiments.

There are numerous directions along which the present
work can be naturally extended. For example, within the
one-dimensional context, it is natural to seek to relax
the simplifying assumption of the intraspecies scattering
lengths (and by extension the self-interaction coefficients
of the two components, $g_{11}$ and $g_{22}$) being equal.
However, this extension will unfortunately have to come
at the expense of an ansatz with different amplitude, width,
etc. parameters between the components, which will render
the intuitive and explicit analytical results obtained herein
much more tedious.

Another natural extension  is to try to generalize
the ideas presented herein into higher-dimensional
or larger number of component (i.e., spinor) settings.
Especially in the former case, more complicated waveforms
such as crosses and propellers have been predicted
in two-component condensates in two-dimensions \cite{borishector}
and it would be particularly interesting to examine
whether these, as well as more complicated multi-hump
waveforms emerge systematically from the corresponding
mixed state via two-dimensional generalizations of
the bifurcations presented herein. Such studies are
presently in progress and will be presented in future
publications.


\begin{acknowledgments}
The authors acknowledge support from NSF-DMS-0806762.
PGK also acknowledges support from NSF-DMS-0349023 (NSF-CAREER)
and from the Alexander von Humboldt Foundation.
\end{acknowledgments}


\begin{thebibliography}{10}
%
\bibitem{Pethick-book}C. J. Pethick and H. S. Smith,
\emph{Bose-Einstein Condensation in Dilute Gases}, Cambridge
University Press, Cambridge, 2002.

\bibitem{stringari} L.P. Pitaevskii and S. Stringari,
{\it Bose-Einstein Condensation}, Oxford University Press (Oxford, 2003).


\bibitem{Our-book}
P.G. Kevrekidis, D.J. Frantzeskakis, and R. Carretero-Gonz{\'a}lez (eds.),
{\it Emergent nonlinear phenomena in Bose-Einstein condensates. Theory and experiment}
(Springer-Verlag, Berlin, 2008).

\bibitem{kivshar} Yu.S. Kivshar and G.P. Agrawal,
{\it Optical solitons: from fibers to photonic crystals},
Academic Press (San Diego, 2003).


\bibitem{shenoy} T.-L.\ Ho and V.B.\ Shenoy, \newblock Phys.\ Rev.\ Lett.\
\textbf{77}, 3276 (1996); H.\ Pu and N.P.\ Bigelow, \newblock Phys.\ Rev.\
Lett.\ \textbf{80}, 1130 (1998).

\bibitem{esry} B.D.\ Esry \textit{et al.}, \newblock Phys.\ Rev.\ Lett.\
\textbf{78}, 3594 (1997).

\bibitem{excit} Th.\ Busch \textit{et al.}, \newblock Phys.\ Rev.\ A \textbf{
56}, 2978 (1997); R.\ Graham and D.\ Walls, \newblock Phys.\ Rev.\ A \textbf{
57}, 484 (1998); H.\ Pu and N.P.\ Bigelow, \newblock Phys.\ Rev.\ Lett.\
\textbf{80}, 1134 (1998); B.D.\ Esry and C.H.\ Greene, \newblock Phys.\
Rev.\ A \textbf{57}, 1265 (1998).


\bibitem{Marek} M.\ Trippenbach,
K.\ Goral, K.\ Rzazewski, B.\ Malomed, and Y.B.\ Band,
J.\ Phys.\ B \textbf{33}, 4017 (2000).

\bibitem{boris2} I.M. Merhasin, B. A. Malomed, R. Driben,
J. Phys. B: At. Mol. Opt. Phys. \textbf{38}, 877-892 (2005).


\bibitem{boris3} I.M. Merhasin, B.A. Malomed and R. Driben,
Phys. Scripta {\bf T116}, 18 (2006).

\bibitem{healt} S.\ Coen and M.\ Haelterman, Phys.\ Rev.\ Lett.\
\textbf{87}, 140401 (2001).



\bibitem{obsantos} P. \"{O}hberg and L.\ Santos, \newblock Phys.\ Rev.\
Lett.\ \textbf{86}, 2918 (2001).


\bibitem{anglin} Th.\ Busch and J.R.\ Anglin, \newblock Phys.\ Rev.\ Lett.\
\textbf{87}, 010401 (2001).

\bibitem{sengstock} C. Becker, S. Stellmer, S. Soltan-Panahi,
S. Dorcher, M. Baumert, E.-M. Richter, J. Kronjager, K. Bongs and K.
Sengstock,
Nature Phys. {\bf 4}, 496 (2008).

\bibitem{ueda_review} K. Kasamatsu, M. Tsubota and M. Ueda,
Int. J. Mod. Phys. B {\bf 19}, 1835 (2005).

\bibitem{decon} B.\ Deconinck, J.N. Kutz, M.S. Patterson
and B.W. Warner, \newblock J.\ Phys.\ A
\textbf{36}, 5431 (2003).

\bibitem{myatt} C.J.\ Myatt,
E.A. Burt, R.W. Ghrist, E.A. Cornell, and C.E. Wieman
\newblock Phys.\ Rev.\ Lett.\
\textbf{78}, 586 (1997).

\bibitem{dsh} D.S.\ Hall,
M.R. Matthews, J.R. Ensher, C.E. Wieman, and
E.A. Cornell, \newblock Phys.\ Rev.\ Lett.\
\textbf{81}, 1539 (1998).

\bibitem{nake} D.M.\ Stamper-Kurn,
M. R. Andrews, A. P. Chikkatur, S. Inouye, H.-J. Miesner, J. Stenger, and
W. Ketterle,
\newblock
Phys.\ Rev.\ Lett.\ \textbf{80}, 2027 (1998).

\bibitem{KRb} G.\ Modugno, G. Ferrari,  G. Roati,
R.J. Brecha,  A. Simoni,  M. Inguscio, \newblock Science \textbf{294},
1320 (2001).

\bibitem{LiCs} M.\ Mudrich, S. Kraft, K. Singer, R. Grimm, A. Mosk, and
M. Weidem{\"u}ller, \newblock Phys.\ Rev.\ Lett.\
\textbf{88}, 253001 (2002).

\bibitem{cornell} V. Schweikhard, I. Coddington, P. Engels, S. Tung
and E.A. Cornell, Phys. Rev. Lett. {\bf 93}, 210403 (2004).

\bibitem{usdsh} K.M. Mertes, J.W. Merrill, R. Carretero-Gonz{\'a}lez,
D.J. Frantzeskakis, P.G. Kevrekidis and D.S. Hall,
Phys. Rev. Lett. {\bf 99}, 190402 (2007).

\bibitem{wieman}  S.B. Papp, J.M. Pino and C.E. Wieman,
Phys. Rev. Lett. {\bf 101}, 040402 (2008).

\bibitem{tsubota} K. Kasamatsu, and M. Tsubota, Phys. Rev. A
{\bf 74}, 013617  (2006).

\bibitem{karali} G. Karali, P.G. Kevrekidis and N.K. Efremidis,
J. Phys. A: Math. Theor {\bf 42}, 045206 (2009).

\bibitem{usreview} R. Carretero-Gonz{\'a}lez,
D.J. Frantzeskakis, and P.G. Kevrekidis, Nonlinearity
{\bf 21}, R139 (2008).

\bibitem{cahn}
M.-S.~Chang, C.D.~Hamley, M.D.~Barrett, J.A.~Sauer,
K.M.~Fortier, W.~Zhang, L.~You, and M.S.~Chapman,
Phys.\ Rev.\ Lett.\ {\bf 92}, 140403 (2004).


\bibitem{Stenger1998a}
J.~Stenger, S.~Inouye, D.M.~Stamper-Kurn, H.-J.~Miesner, A.P.~Chikkatur, and
W.~Ketterle.
Nature {\bf 396}, 345 (1998).


\bibitem{spindw}
H.E.~Nistazakis, D.J.~Frantzeskakis, P.G.~Kevrekidis,
B.A.~Malomed, R.~Carretero-Gonz\'alez, and A.R.~Bishop, Phys.\ Rev.\ A {\bf 76}, 063603 (2007).

\bibitem{spintext}
A.E.~Leanhardt, Y.~Shin, D.~Kielpinski, D.E.~Pritchard, and W.~Ketterle,
Phys.\ Rev.\ Lett.\ {\bf 90}, 140403 (2003).

\bibitem{wad1a}
J.~Ieda, T.~Miyakawa, and M.~Wadati,
Phys.\ Rev.\ Lett.\ {\bf 93}, 194102 (2004).
\bibitem{wad1b}
J.~Ieda, T.~Miyakawa, and M.~Wadati,
J.\ Phys.\ Soc.\ Jpn.\ {\bf 73}, 2996 (2004).

\bibitem{boris}
L.~Li, Z.~Li, B.A.~Malomed, D.~Mihalache, and W.M.~Liu,
Phys.\ Rev.\ A {\bf 72}, 033611 (2005).

\bibitem{zh}
W.~Zhang, \"{O}.E.~M\"{u}stecaplioglu, and L.~You,
Phys.\ Rev.\ A {\bf 75}, 043601 (2007).

\bibitem{wad2}
M.~Uchiyama, J.~Ieda, and M.~Wadati,
J.\ Phys.\ Soc.\ Jpn.\ {\bf 75}, 064002 (2006).

\bibitem{ofyspin}
B.J.~Dabrowska-W\"{u}ster, E.A.~Ostrovskaya,
T.J.~Alexander, and Y.S.~Kivshar,
Phys.\ Rev.\ A {\bf 75}, 023617 (2007).

\bibitem{ourspin}
H.E.~Nistazakis, D.J.~Frantzeskakis, P.G.~Kevrekidis,
B.A.~Malomed, and R.~Carretero-Gonz\'alez,
Phys. Rev. A {\bf 77}, 033612 (2008).

\bibitem{oberthaler} A. Weller, J.P. Ronzheimer, C. Gross,
J. Esteve, M.K. Oberthaler, D.J. Frantzeskakis,
G. Theocharis, and P.G. Kevrekidis,
Phys. Rev. Lett. {\bf 101}, 130401 (2008).

\bibitem{borishector} B.A. Malomed, H.E. Nistazakis,
D.J. Frantzeskakis, and P.G. Kevrekidis,
Phys. Rev. A {\bf 70}, 043616 (2004).

%

\end{thebibliography}
\end{document}